\documentclass[fleqn,usenatbib]{mnras}
\usepackage[T1]{fontenc}
\usepackage{ae,aecompl}

\usepackage{graphicx}	% Including figure files
\usepackage{amsmath}	% Advanced maths commands
\usepackage{amssymb}	% Extra maths symbols
\usepackage{journals}
\usepackage[dvipsnames]{color}
\usepackage{threeparttable}

\newcommand{\OIII}{[O~{\sc iii}]\ }

\newcommand{\HII}{H~{\sc ii}\ }

\newcommand{\HI}{H~{\sc i}\ }
\newcommand{\Ha}{H$\alpha$\ }
\newcommand{\kms}{\,\mbox{km}\,\mbox{s}^{-1}}

\newcommand{\SIIHa}{[S~{\sc ii}]/H$\alpha$}
\newcommand{\NIIHa}{[N~{\sc ii}]/H$\alpha$}
\newcommand{\OIIIHb}{[O~{\sc iii}]/H$\beta$}
\newcommand{\sunn}{$_{\odot}$}
\newcounter{qub}
\newcommand{\qq}{\addtocounter{qub}{1}\arabic{qub}}

\usepackage{etoolbox}
\makeatletter
\newcount\c@additionalboxlevel
\newcount\c@maxboxlevel
\patchcmd\@combinedblfloats{\box\@outputbox}{%
	\stepcounter{additionalboxlevel}%
	\box\@outputbox
}{}{\errmessage{\noexpand\@combinedblfloats could not be patched}}

\AtBeginShipout{%
	\ifnum\value{additionalboxlevel}>\value{maxboxlevel}%
	\typeout{Warning: maxboxlevel might be too small, increase to %
		\the\value{additionalboxlevel}%
	}%
	\fi
	\@whilenum\value{additionalboxlevel}<\value{maxboxlevel}\do{%
		\typeout{* Additional boxing of page `\thepage'}%
		\setbox\AtBeginShipoutBox=\hbox{\copy\AtBeginShipoutBox}%
		\stepcounter{additionalboxlevel}%
	}%
	\setcounter{additionalboxlevel}{0}%
}
\makeatother
\title[Gas accretion in voids: NGC 428 galaxy]{Search for gas accretion imprints in voids: I. Sample selection and results for NGC 428}

\author[Egorova et al.]{
Evgeniya S. Egorova,$^{1,2}$\thanks{E-mail: eshaldenkova@gmail.com}
Alexei V. Moiseev,$^{1,2,3}$
Oleg V. Egorov$^{1,2}$
\\
$^{1}$ Lomonosov Moscow State University, Sternberg Astronomical Institute,	Universitetsky pr. 13, Moscow 119234, Russia \\
$^{2}$ Special Astrophysical Observatory, Russian Academy of Sciences, Nizhny Arkhyz 369167, Russia \\
$^{3}$ Space Research Institute, Russian Academy of Sciences, Profsoyuznaya ul. 84/32, Moscow 117997, Russia
}

\date{Accepted XXX. Received YYY; in original form ZZZ}

\pubyear{2018}

\begin{document}
\label{firstpage}
\pagerange{\pageref{firstpage}--\pageref{lastpage}}
\maketitle


\begin{abstract}
We present the first results of a project aimed at searching for gas accretion events and interactions between late-type galaxies in the void environment. The project is based on long-slit spectroscopic and scanning Fabry-Perot interferometer observations performed with the SCORPIO and SCORPIO-2 multimode instruments at the Russian 6-m telescope, as well as archival multiwavelength photometric data. In the first paper of the series we describe the project and present a sample of 18 void galaxies with oxygen abundances that fall below the reference `metallicity-luminosity' relation, or with possible signs of recent external accretion in their optical morphology. To demonstrate our approach, we considered the brightest sample galaxy NGC 428, a late-type barred spiral with several morphological peculiarities. We analysed the radial metallicity distribution, the ionized gas line-of-sight velocity and velocity dispersion maps together with WISE and SDSS images. Despite its very perturbed morphology, the velocity field of ionized gas in NGC~428 is well described by pure circular rotation in a thin flat disc with streaming motions in the central bar. We also found some local non-circular gas motions clearly related to stellar feedback processes. At the same time, we revealed a circumnuclear inclined disc in NGC~428 and a region with significant residual velocities that could be considered as a result of a recent ($<0.5$ Gyr) accretion event. The observed oxygen abundance gradient does not contradict this conclusion.

\end{abstract}


%%%%%%%%%%%%%%%%%%%%%%%%%%%%%%%%%%%%%%%%%%%%%%%%%%

%%%%%%%%%%%%%%%%% BODY OF PAPER %%%%%%%%%%%%%%%%%%

\section{Introduction}

With the current star formation rate (SFR), the gas reservoirs in present-day galaxies should be exhausted in 1--2~Gyrs \citep{Bigiel08,Leroy2008,Bigiel2011,Leroy2013}. Since this value is significantly shorter than the Hubble time, a gas supply is needed to sustain star formation \citep{Lilly2013}. In different papers, several mechanisms of galaxy growth and gas replenishment are discussed, including mergers \citep[e.g.,][]{LHuillier2012,DiMatteo2008}, galactic fountains \citep[e.g.,][]{Fraternali2006,Fraternali2008,Marinacci2010}, and cold gas accretion from filaments \citep{Semelin2005,Keres2005,Dekel2006,Dekel2009}. A number of simulations have shown that smooth accretion from filaments dominates over mergers \citep{LHuillier2012,Wang2011,vandeVoort2011}.

%About voids, why their are suitable to look for the accretion.

The aim of our work is to study the interactions and possible evidence of gas accretion (via mergers or cold gas accretion from filaments) in voids. Due to a low rate of interactions and mergers in voids compared to dense environments, voids are well-suited for this purpose -- it is easier to distinguish between various scenarios of interaction/accretion and between different sources of accreting gas.

Several recent works are dedicated to the study of samples of galaxies residing in the Lynx-Cancer (\citealt{LC1}, \citealt*{LC2}, \citealt*{LC6}, \citealt*{LC7}) and Eridanus \citep*{Eridanus} voids. It was found that these void galaxies have, on average, substantially lower metallicities than those in the control sample of similar-type Local Volume galaxies \citep{Berg12}. This was treated as evidence of a slower evolution of galaxies in voids. A small subgroup of objects demonstrates very low gas metallicity, a high gas mass fraction ($\sim94-99$ per cent), and atypical periphery colours corresponding to the time elapsed since the onset of star formation $\sim1-3.5$ Gyr \citep*{LC4}. Such properties are consistent with the unevolved state of these galaxies, which are among the least-massive in voids.

%On the other hand, the substantial fraction of intermediate- luminosity galaxies in voids also reveal lower metallicity for their luminosity in comparison with the `reference' relation \citep*{LC7}. 
Low-mass galaxies are not the only ones to exhibit metallicities lower than expected for their luminosity in comparison with the `reference' relation \citep*{LC7}.
In the case of intermediate-luminosity void objects, the reduced metallicity could be due to tidal interactions \citep[e.g.,][]{Ekta10}, mergers \citep{Bekki2008,Montuori2010}, or accretion of metal-poor gas from the filaments \citep{SA2014}. Such events may also lead to an increase in the SFR.

Although a detailed study of interactions and various gas accretion scenarios (e.g., accretion from companion galaxies or via mergers) in low-density environments is in itself of great interest, the most intriguing part is a search for cold gas accretion from filaments. Cold accretion from filaments is assumed to occur at high redshifts, however, in the case of low-mass galaxies, it may also take place at the present time \citep{Birnboim2003}. % Cold accretion from filaments may take place at high redshifts, or in recent times, in the case of low-mass galaxies \citep{Birnboim2003}. 
Simulations by \citet{Aragon13} have also shown that haloes in voids could accrete gas from the cosmic web in a steady and coherent way for long periods of time.  

A good observational example is the well-known  Hoag's object. A study of its morphology, kinematics, and stellar population properties revealed evidence that the massive star-forming ring is a result of prolonged cold gas accretion from a filament onto an isolated galaxy in a relatively low density environment  \citep{Finkelman2011}. Moreover, a possible remnant of this filament appeared in HI observations \citep{Brosch2013}.

Our study is aimed at searching for signs of gas accretion and interactions between void galaxies. The main idea is to combine the data on ionized-gas kinematics, optical and NIR morphology, and chemical abundances. We are looking for morphological peculiarities (e.g., tidal features, star formation on the periphery of a disc, asymmetric and disturbed appearance), and for misalignments between gas kinematics and optical morphology. In the latter case, 3D spectral data, i.e., mapping gas kinematics parameters (line-of-sight velocities and velocity dispersion) are essential. For this purpose, we use a scanning Fabry-Perot interferometer (FPI), because this device provides the best combination of a large field-of-view, high spectral resolution, and emission-line detection limit as compared with other integral-field spectrographs.  Using these data together with the information on the chemical abundance distribution, we can make assumptions about the source of the disturbances, or about the source of external gas in the case of accretion events.

%The possible observational example could be the UGC 3672A studied by \cite{Chengalur17} and residing near the void centre. It reveal extremely low metallicity and colours consistent with young stellar population on the periphery. UGC 3672 system could be interpreted as triplet with its members lying along filamentary structure, and the unusual properties of UGC 3672A component may be related with inflow along the filament to galaxy pair. In such a case star formation in UGC 3672A could be triggered by tidal disturbance from main system <~1 Gyrs ago.
%The accretion of the metal-poor gas may lead to the metallicty drops and SFR enhancements (see \cite{SA2014} and references therein).

%We compiled the sample of galaxies residing in the Lynx-Cancer and Eridanus voids. We selected intermediate luminosity galaxies with reduced metallicity for their luminosity (comparing with galaxies in more dense environment -- the sample from \citealt{Berg12}) and/or with signs of disturbances -- bright SF regions on the periphery, tidal features etc. 

This paper is the first in a series dedicated to  searching for  gas accretion and interactions in voids. Here we present our current sample and  an investigation of the brightest galaxy in that sample -- NGC 428.  The case of this galaxy presents a good illustration of our data-analysing techniques, because despite its peculiar optical morphology, the properties of ionized gas on large radial scales follow the relations typical of ordinary late-type disc galaxies. However, a detailed analysis revealed some small-scale peculiarities in the gas kinematics and excitation, which could be related to a recent accretion event. The study of the remaining targets in the sample, including several objects with very strong misalignments between their morphology and kinematics, will be presented in the upcoming papers.

The paper is organised as follows. In Section \ref{sample} we briefly describe the selection criteria and the sample of void galaxies. In Section \ref{system} we present an overview of the NGC 428 system. In Section \ref{data} we describe our observations and data reduction. In Section \ref{results} we show the results of observations and describe the performed analysis. In Section \ref{discussion} we discuss the obtained results, and Section \ref{conclusions} summarises our conclusions.

\section{Sample overview}
\label{sample}

The sample consists of intermediate luminosity galaxies from the Lynx-Cancer (\citealt{LC1}, \citealt{LC2}, \citealt{LC6}, \citealt{LC7}) and Eridanus \citep{Eridanus}
voids. We selected %intermediate luminosity 
galaxies which have lower metallicities than expected for their luminosity (compared to a sample of galaxies in a denser environment from \citealt{Berg12}) and/or with signs of disturbances -- bright star-forming regions on the periphery, tidal/lopsided features and other asymmetric features. 
The sample includes both isolated galaxies and systems at different stages of interaction. 

\begin{figure}
\includegraphics[angle=0,width=\columnwidth]{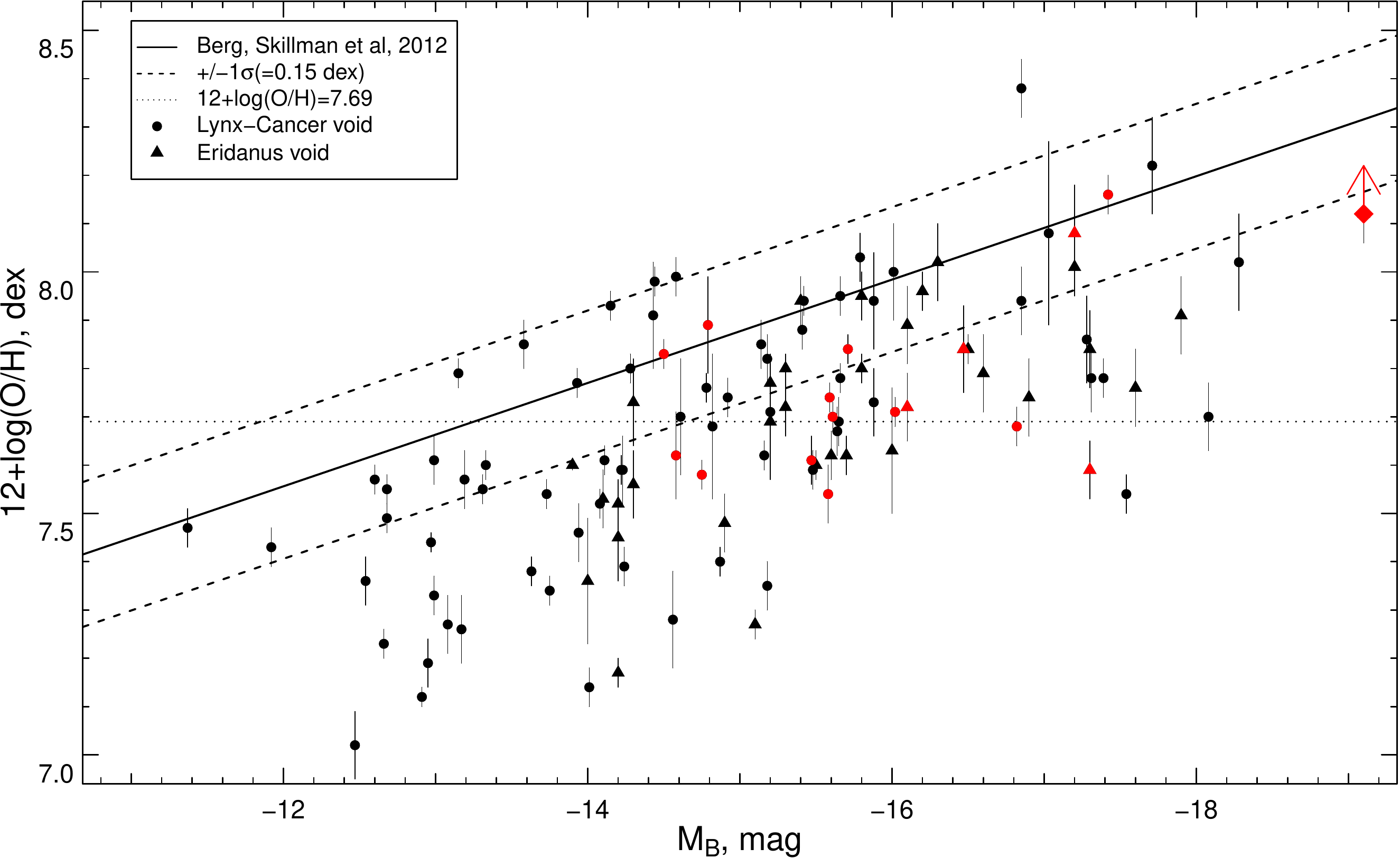}
    \caption{Relation between $\log$(O/H) and the absolute blue magnitude $M_{\rm B}$ for the Lynx-Cancer (circles) and Eridanus (triangles) void galaxies. The red symbols denote our sample galaxies. The red diamond denotes the NGC 428 galaxy.  The red arrow points to the value obtained in our study (see Table~\ref{tab:oh}). 
The solid line shows the linear regression
for the control sample from the Local Volume by \citet{Berg12}.
Two dashed lines on both sides of the reference line show the r.m.s. deviation of their sample from the linear regression (0.15 dex). 
The horizontal dotted black line marks the value 12+$\log$(O/H)=7.69 which corresponds to Z\sunn/10 for Z\sunn\ from \citet{AllendePrieto2001} and \citet{Asplund09}.}
    \label{fig:sample}
\end{figure}

%Almost all galaxies reveal signs of disturbances -- bright SF regions on the periphery, tidal features etc. Also most of the sample objects have lower metallicity for their luminosity in comparison with the `reference' relation obtained for the Local Volume galaxies \citep{Berg12}. 
The `metallicity--luminosity' relation for both Lynx-Cancer and Eridanus void galaxies is presented in Fig.\ref{fig:sample}. The galaxies from our sample are marked with the red symbols. The 12+log(O/H) values are taken from the papers by \citet{LC2,LC7} and \citet{Eridanus}. 
Table~\ref{tab:sample} presents the main parameters of the sample galaxies:

\begin{description}
\item[$\bullet$] Column 1: Common name.
\item[$\bullet$] Column 2: Epoch J2000 R.A.
\item[$\bullet$] Column 3: Epoch J2000 Declination.
\item[$\bullet$] Column 4: Heliocentric velocity. 
\item[$\bullet$] Column 5: Distance D, in Mpc, adopted as $V_{\rm LG}/H$ (assuming the Hubble constant $H=\mathrm{73\ km\ s^{-1}\ Mpc^{-1}}$). Here $V_{\rm LG}$ is the recession velocity in the Local Group coordinate system. The sources are given by letters in superscript for the respective values. %For Lynx-Cancer galaxies also the correction for the large peculiar velocity was made as in \citet{LC1}} 
\item[$\bullet$] Column 6: The apparent total $B$-band magnitude $B_{\rm tot}$ (from NED or from the literature). %The sources of $B$-mag are given by a letter in the superscript for the respective values. 
\item[$\bullet$] Column 7: Absolute $B$-band magnitude $M_{\rm B}$ calculated using $B_{\rm tot}$ corrected for the Galactic foreground extinction $A_{\rm B}$ (from NED, following \citealt{Schlafly2011}).
\item[$\bullet$] Column 8: the value of 12+log(O/H) adopted from \citet{LC7} and \citet{Eridanus}.
\item[$\bullet$] Column 9: Comments on the environment.
%Isolated / with satelites / interacting ???}
\end{description}

For Ark~18, MCG~-01-03-027, Mrk~965, MCG~-01-03-072 and NGC~428, all values are taken from tables 1 and 3 in \citet{Eridanus}; for the remaining galaxies -- from tables 2 and 3 in \citet{LC7}.

In total, we selected 18 galaxies with absolute magnitudes $M_{\rm B}$ ranging from $-19.1^m$ to $-14.5^m$. %Among them ??? galaxies show significantly reduced metallicity for their luminosity. %??? galaxies from our sample exhibit a peculiar morphology in the outer parts of the disc. 
%While for most sample galaxies there were observed any satellite or even nearby companion with a similar mass (which galaxies exactly, if any??), ??? galaxies are believed to be very isolated (which exactly). 
Ten objects were included in the samples of nearby isolated galaxies by \citet{Karachentseva2010,Karachentsev2011,Karachentsev2013}, while others are mainly part of systems with different stages of interaction. Almost all our sample galaxies exhibit a peculiar morphology, although more than half of them are considered to be isolated with no known companions. Further in this paper, we will focus on the brightest galaxy in our sample -- NGC~428 -- which demonstrates both peculiar morphology and low metallicity according to the data in the literature.

\begin{table*}
\centering
\caption{Sample galaxies}
\label{tab:sample}
\begin{tabular}{rlllllllll} 
\hline
\multicolumn{1}{c|}{\#}                 &
\multicolumn{1}{l|}{Name} & 
\multicolumn{2}{c|}{Coordinates (J2000)}& 
\multicolumn{1}{c|}{$V_{hel}$,} & 
\multicolumn{1}{c|}{$D,$} & 
\multicolumn{1}{c|}{$B_{tot}$,} & 
\multicolumn{1}{c|}{$M_B$,} & 
\multicolumn{1}{c|}{12+log(O/H),}  &
\multicolumn{1}{c|}{Comments}  \\
&
\multicolumn{1}{l|}{} & 
\multicolumn{2}{c|}{of objects} & 
\multicolumn{1}{c|}{km/s} & 
\multicolumn{1}{c|}{Mpc} & 
\multicolumn{1}{c|}{mag} & 
\multicolumn{1}{c|}{mag} & 
\multicolumn{1}{c|}{dex} &
\multicolumn{1}{c|}{}  \\ 
&
\multicolumn{1}{l|}{(1)} & 
\multicolumn{1}{c|}{(2)} & 
\multicolumn{1}{c|}{(3)} & 
\multicolumn{1}{c|}{(4)} & 
\multicolumn{1}{c|}{(5)} & 
\multicolumn{1}{c|}{(6)} & 
\multicolumn{1}{c|}{(7)} &
\multicolumn{1}{c|}{(8)} &
\multicolumn{1}{c|}{(9)}  \\ 
\hline 
\qq&Ark 18         & 00 51 59.62 & -00 29 12.2 & 1621 &24.1$^a$& 14.76 & -17.2 & 8.08$\pm$0.10 & isolated$^1$ \\
\qq&MCG -01-03-027 & 00 52 17.23 & -03 57 59.8 & 1412 &21.1$^a$& 15.69 & -16.1 & 7.72$\pm$0.07 & isolated$^1$ \\
\qq&Mrk 965        & 00 57 28.75 & -04 09 34.0 & 2713 &38.8$^a$& 15.81 & -17.3 & 7.59$\pm$0.06 & \\
\qq&MCG -01-03-072 & 01 02 22.91 & -04 30 30.9 & 1764 &25.8$^a$& 15.69 & -16.5 & 7.84$\pm$0.09 & \\
\qq&NGC 428        & 01 12 55.71 & +00 58 53.6 & 1152 &17.6$^a$& 12.16 & -19.1 & 8.12$\pm$0.06 & in group \\
\qq&UGC 3476       & 06 30 29.22 & +33 18 07.2 & 469 &9.8$^b$& 14.96 & -16.0 & 7.71$\pm$0.03 & isolated$^2$ \\
\qq&UGC 3860       & 07 28 17.20 & +40 46 13.0 & 354 &7.8$^b$& 15.21 & -14.5 & 7.83$\pm$0.03 & isolated$^2$ \\
\qq&UGC 3966       & 07 41 26.00 & +40 06 44.0 & 361 &8.6$^b$& 15.32 & -14.6 & 7.62$\pm$0.09 & isolated$^2$ \\
\qq&UGC 4115       & 07 57 01.80 & +14 23 27.0 & 341 &7.7$^b$& 14.81 & -14.8 & 7.61$\pm$0.06 & isolated$^1$ \\
\qq&UGC 4117       & 07 57 25.98 & +35 56 21.0 & 773 &14.1$^b$& 14.81 & -15.6 & 7.73$\pm$0.03 & isolated$^1$ \\
\qq&NGC 2552       & 08 19 20.14 & +50 00 25.2 & 524 &11.1$^b$& 13.01 & -17.4 & 8.16$\pm$0.04 & \\
\qq&KUG 0934+277   & 09 37 47.65 & +27 33 57.7 & 1588&25.1$^b$& 16.50 & -15.6 & 7.55$\pm$0.08 & isolated$^3$\\
\qq&IC 559         & 09 44 43.82 & +09 36 57.5 & 541 &9.4$^b$& 14.98 & -14.9 & 7.89$\pm$0.10 & isolated$^2$ \\
\qq&Mrk 407        & 09 47 47.60 & +39 05 03.0 & 1589&25.2$^b$& 15.27 & -16.8 & 7.68$\pm$0.05 & \\
\qq&UGC 5272       & 09 50 22.40 & +31 29 16.0 & 520 &10.3$^b$& 14.45 & -15.7 & 7.84$\pm$0.03 & in pair \\
\qq&UGC 5288       & 09 51 16.77 & +07 49 47.9 & 556 &9.5$^b$& 14.42 & -15.6 & 7.66$\pm$0.05 & isolated$^2$ \\
\qq&IC 2520        & 09 56 20.12 & +27 13 39.3 & 1243&19.9$^c$& 14.27 & -17.3 & -- & \\
\qq&UGC 5464       & 10 08 07.70 & +29 32 34.4 & 1003&16.9$^b$& 15.77 & -15.5 & 7.61$\pm$0.06 & \\
\hline
\end{tabular}
\begin{tablenotes}
%\item * the value 12+log(O/H) for NGC 428 from paper Kniazev et al, 2018 (submitted to MNRAS), differs from that derived in our study 
\item $^a$ from NASA/IPAC Extragalactic Database (NED)
\item $^b$ from \citet{LC7}
\item $^c$ from \citet{LC6}
\item $^1$ included in the "Local Orphan Galaxies" sample by \citet{Karachentsev2011};
\item $^2$ according to \citet{Karachentsev2013};
\item $^3$ according to \citet{Karachentseva2010}.
\end{tablenotes}
\end{table*}

\section{NGC\,428: main properties and environment}
\label{system}

\begin{figure}
\includegraphics[width=\columnwidth]{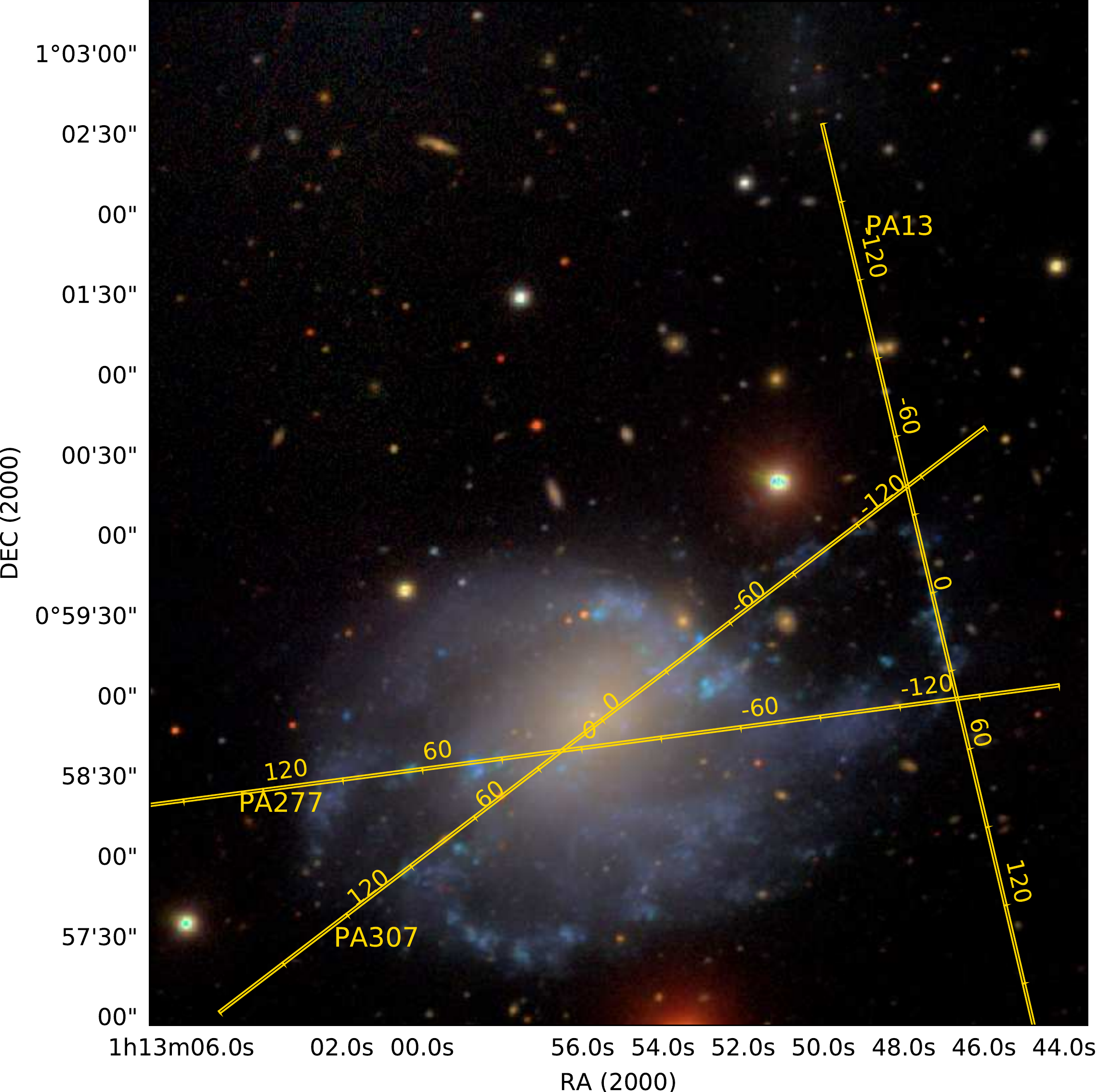}
    \caption{SDSS Stripe82 $g$ (blue), $r$ (green) and $i$ (red) band image of NGC 428 and its vicinity. Positions and labels of the slits used during spectroscopic observations are shown by yellow lines.}
    \label{fig:slits}
\end{figure}

NGC~428 is a late-type spiral galaxy (SABm) and a dominant member of a loose group. Its main properties are presented in Table~\ref{tab:sample}. The galaxy looks quite perturbed, with strongly asymmetric spiral arms and many bright star-forming regions distributed irregularly throughout the disc \citep{Hutchings1996,Eskridge2002}. The most intriguing feature is a huge ($\sim7$~kpc in  diameter) blue ring of  star-forming knots  in the northwestern outskirts of the disc that is clearly visible in the optical SDSS and SDSS Stripe82 images (Fig.~\ref{fig:slits}). 
% \red{ Is it Stripe82 or not???}

Other galaxies in the  NGC~428 group are: UGC~772, MCG~+00-04-049, LEDA~135629 (=J0112+0102). J0112+0102 is a low surface brightness galaxy $\sim 3 \arcmin$ north of NCG 428 (see Fig.~\ref{fig:slits}) with no $12+\log(\mathrm{O/H})$ measurements. UGC~772 and MCG~+00-04-049 are metal-poor dwarf galaxies with $12+\log(\mathrm{O/H}) = 7.15-7.32$ (for several knots, \citealt{Izotov2012}) and $7.56\pm0.07$ \citep{Eridanus}, respectively. For UGC 772, \citet{Moiseev2010} studied the H$\alpha$-line ionized-gas kinematics; \citet{Ekta2008} also presented the results of \HI observations with GMRT. In both papers, the authors concluded that this galaxy had experienced a recent  merger event.

\HI observations presented in \citet{Smoker1996} reveal an \HI tail to the south of NGC~428. The authors discussed that this might be a result of tidal interaction between the group members. Another suggested possibility is that NGC~428 is a result of a merger; in this case, J0112+0102 may have been formed from the merger remnants.
%Recently one more metal-poor component was discovered (paper in preparation).

\begin{table*}
\centering
\caption{Observation log}
\label{tab:Obs}
\begin{tabular}{llllllll} 
\hline
{Data set} & {Date of obs.} & {$T_{exp}$, s}& {$FOV$} & {$\arcsec/px$} & {Seeing, $\arcsec$} & {sp. range} & {$\delta\lambda$, \AA} \\
{(1)} & {(2)} & {(3)} & {(4)} & {(5)} & {(6)} & {(7)} & {(8)}  \\  
\hline 
LS PA=277 & 2015.09.20 & 1800 & {$1\arcsec\times6.1\arcmin$} &  0.36 & 2.0 & 3500--7500 & 13  \\
LS PA=307 & 2015.09.20 & 3600 & {$1\arcsec\times6.1\arcmin$} & 0.36 & 2.0  & 3500--7500 & 13\\
LS PA=13 & 2015.09.20 & 3000 & {$1\arcsec\times6.1\arcmin$} & 0.36 & 2.0 & 3500--7500 & 13 \\
FPI & 2015.11.04 & $40\times90$ & {$6.1\arcmin\times6.1\arcmin$} & 0.71 & 1.8   & 8.8\AA\, around H$\alpha$ & 0.48 ($22 \kms$) \\ %1.1-1.8 seeing before data reduction
\hline
\end{tabular}
\end{table*}

NGC~428 is located in a rather peculiar low-density environment together with the other group members, revealing unusual properties. It is a good candidate for the study of accretion and interactions in voids and clearly deserves a more thorough investigation.

The adopted distance to the galaxy is 17.6 Mpc according to \citet{Eridanus}, calculated as $V_{\rm LG}/H$ (assuming Hubble constant $H=73\ \mathrm{km\ s^{-1}\ Mpc^{-1}}$), where $V_{\rm LG}$ is the recession velocity in the Local Group coordinate system. At the adopted distance the scale is 85~pc~arcsec$^{-1}$.

\section{Observations and data reduction}
\label{data}

\subsection{Long-slit spectroscopic observations}

Spectral data were obtained with the SCORPIO multimode spectrograph \citep{SCORPIO} at the  6-m Russian telescope (BTA) of SAO RAS. We used a VPHG550G grism covering the range of 3500 to 7500~\AA\, with a 1.0 arcsec slit width, which provides a typical spectral resolution of 13~\AA\, as estimated from the \textit{FWHM} of air-glow emission lines. We obtained three spectra for different slit positions shown in Fig.~\ref{fig:slits}. The observation log is presented in Table~\ref{tab:Obs}, where $T_{exp}$ is the exposure time, $FOV$ is the field of view,  $''/px$ is the pixel size in the final images, and  $\delta\lambda$ is the final spectral resolution. The quantities in the Seeing column in the table correspond to the final angular resolution in the summed exposures.

The data were reduced in a standard way using the \textsc{idl}-based pipeline developed for the SCORPIO data. The main steps of the data reduction process include bias subtraction,
line curvature and flat-field corrections, wavelength calibration, and air-glow line subtraction. The spectra were calibrated to the wavelength scale using the He-Ne-Ar lamp reference spectrum obtained during the observations. One of the spectrophotometric standards (either AGK+81d266 or BD+25d4655) was observed at a close zenith distance immediately before or after the object, and was used for the absolute intensity scale calibration.

To measure the emission line fluxes, we used our own software working in \textsc{idl} environment and based on the \textsc{mpfit} \citep{mpfit} routine. In order to increase the signal-to-noise ratio, all spectra were binned along the slit using 3 px ($\sim1$ arcsec) bins. Afterwards, we performed Gaussian fitting to measure the integrated line fluxes in each studied region. To subtract the spectrum of an underlying stellar population, we performed modelling using the \textsc{ULySS}\footnote{\url{http://ulyss.univ-lyon1.fr}} package \citep{Koleva2009}. To estimate the final uncertainties of the measured line fluxes, we quadratically added the errors propagated through all data-reduction steps to the uncertainties returned by \textsc{mpfit}.

All the measured fluxes used in this paper are reddening corrected. The colour excess $E(B-V)$ was derived from the observed Balmer decrement, and the \citet*{Cardelli1989} curve parametrized by \citet{Fitzpatrick1999} was used to perform the reddening correction.
In this paper, we use the following abbreviations for the emission-line flux ratios: \SIIHa\, is F([S~{\sc ii}] 6717,6731\AA)/F(H$\alpha$); \NIIHa\, is F([N~{\sc ii}] 6584\AA)/F(H$\alpha$); \OIIIHb\, is F([O~{\sc iii}] 5007\AA)/F(H$\beta$).

\subsection{FPI observations}

Observations were carried out in the prime focus of
the 6-m Russian telescope (BTA) with a scanning Fabry-Perot interferometer (FPI) mounted inside the SCORPIO-2 multimode focal reducer \citep{SCORPIO2}. 
During the scanning process, we have consecutively obtained 40 interferograms at different gaps between the FPI plates. The galaxy  was exposed at two position angles in order to remove the parasitic ghosts following  \cite{Moiseev2008}. The details of the observations are presented in Table~\ref{tab:Obs}.

The data were reduced using a software
package running in the \textsc{idl} environment. For a detailed description of the data reduction algorithms, see \citet{Moiseev2002,Moiseev2015}. After initial reduction, the observed data were combined into data cubes, where each pixel in the field of view contains a 40-channel spectrum around the redshifted \Ha emission line.

H$\alpha$-line profiles were analysed using one-component Voigt fitting \citep{Moiseev2008} which yields the flux, line-of-sight velocity, and velocity dispersion corrected for instrumental broadening for each component. Several areas exhibiting asymmetric H$\alpha$-line profiles were analysed separately by multicomponent Voigt fitting.

\begin{figure*}
		\includegraphics[width=\linewidth]{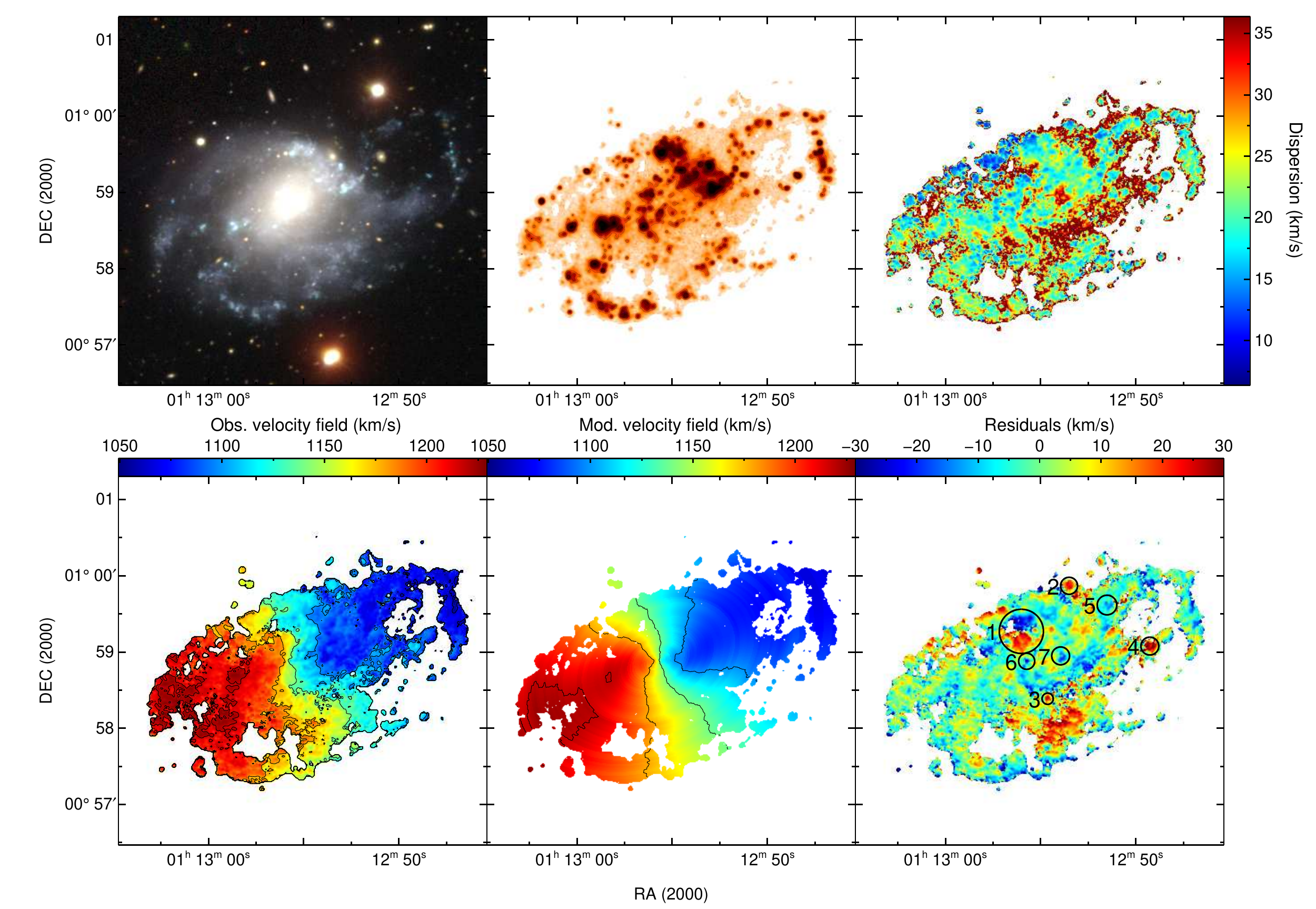}
    \caption{\textbf{Top row:} the coloured SDSS Stripe82 gri-image (left-hand panel); the map of $H\alpha$ fluxes (middle panel); the map of the ionized-gas velocity dispersions (right-hand panel). \textbf{Bottom row:} the observed ionized-gas velocity field (left-hand panel); the velocity field obtained with a tilted-ring model (middle panel); the map of residuals after subtracting the model velocity field from the observed one (right-hand panel). The circles in the bottom right-hand panel mark the regions with significant residual velocities that are shown in Fig.~\ref{fig:resid_profile} and discussed in Section~\ref{sec:resid}. }
    \label{fig:FPI}
\end{figure*}

The resulting maps with a final angular resolution of 2\farcs1 are shown in Fig.~\ref{fig:FPI}. Thanks to the large aperture and fast focal ratio of BTA/SCORPIO-2, we were able to obtain significantly deeper data compared to the previous FPI observations by \citet{GHASP_IV} and \citet{ErrozFerrer2015}.  Namely, we have mapped the velocity and distribution of diffuse gas between the bright \HII regions.

\subsection{Other observed data used}

To perform the surface photometry of NGC~428 we used 3.4$\mu$m (W1 band) archival images from the Wide-field Infrared Survey Explorer \citep[WISE;][]{WISE}. The AllWISE data release combines the data from the cryogenic and post-cryogenic \citep{Mainzer2011} survey phases\footnote{\url{http://irsa.ipac.caltech.edu/cgi-bin/Gator/nph-scan?submit=Select\&projshort=WISE}}.

Additionally, we also used the 2MASS \citep{2MASS} \textit{J}-band and SDSS 
Stripe82 \citep{Stripe82} \textit{gri}-band images for aperture photometry.
In the SDSS Stripe82 images, a ghost from a bright star was projected onto the galaxy. We downloaded individual SDSS Stripe82 exposures and masked the ghost in all the frames where it appeared. After that, all the individual images were combined into a mosaic.

\section{Results}
\label{results}

\subsection{Photometric structure and global ionized-gas kinematics}
\label{sec:kin}

The peculiar optical morphology of the galaxy was already mentioned in the previous studies \citep[e.g.,][]{Eskridge2002}. Based on the coloured SDSS  images shown in Figs.\ref{fig:slits} and \ref{fig:FPI}, we can propose  two  contradictory interpretations  of the NGC~428 structure:
\begin{enumerate}
\item  A classical dwarf disc galaxy with a large red elliptical bar related to two tightly-bound strongly asymmetric spiral arms. This is the generally accepted  view of the galaxy, presented in, e.g., \citet{Smoker1996} and \citet{GHASP_IV}.
\item  A multispin system with two misaligned components: a central early-type red spheroid surrounded by a stellar-gaseous ring accreted from matter with different angular momentum orientations. Both components are moderately inclined to the line-of-sight similar to some polar-ring galaxy candidates like SPRC-125 or SPRC-241 \citep[Sloan-based Polar Ring Catalogue,][]{Moiseev2011}. Spiral and quasi-spiral structures are also known among confirmed polar ring galaxies \citep{Iodice2002,Brosch2010}.
\end{enumerate}

\begin{figure}
	\includegraphics[angle=-90,width=\columnwidth]{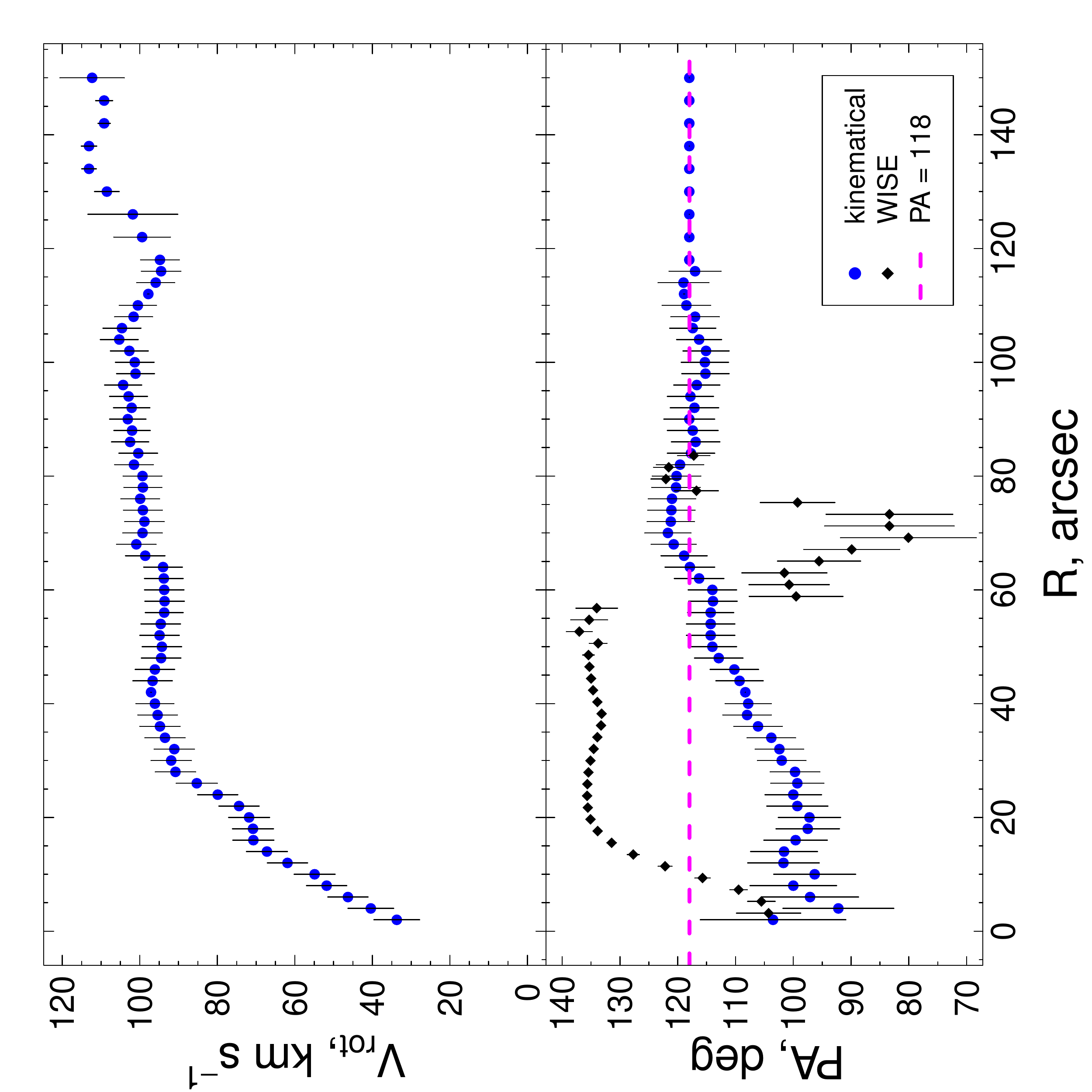}
    \caption{\textbf{Top:} Rotation curve for NGC 428 obtained from the observed ionized gas velocity field using tilted-ring analysis.
    \textbf{Bottom:} The distribution of position angles along the major axis. The values obtained by tilted-ring analysis of the observed velocity field are marked by blue circles, the values derived from isophotal analysis of WISE data are marked by black diamonds. The line-of-nodes position angle $PA_0$ is shown by the purple dashed line.}
    \label{fig:model}
\end{figure}

Analysing the two-dimensional velocity field allows us to distinguish between 
the above possibilities based on the position angles of the kinematic major 
axis $PA_{kin}$, determined as a major axis of the orbits projected onto the sky 
plane. In the case of circular motions in a polar or warped disc,  $PA_{kin}$ is in good agreement with the position angle of the elliptical isophotes $PA_{phot}$. However, in the case of a bar, gas streaming motions lead to  $PA_{kin}$ being turned from the disc's line of nodes $PA_0$ in the direction opposite to the major axis of the inner isophotes ($PA_{phot}$). The deviation of $PA_{kin}$  from  $PA_0$ depends on the bar strength and the angle between the disc's line of nodes and the bar axis \citep[see, e.g., simulations in][]{Moiseev2000}. The description of the $PA_{kin}$ and $PA_{phot}$ comparison technique and references to the previous papers are presented in \citet{Moiseev2000}.

The observed \Ha velocity field   was analysed by means   of  the classical tilted-ring method \citep{Begeman1989A} adapted for the study of ionized-gas velocities in dwarf galaxies from \citet{Moiseev2014}.  The rotation centre, determined as the velocity-field symmetry centre, matches very well (better that 1 px)  the centre of  continuum isophotes. We then fixed the location of the centre and split the velocity field into narrow elliptical rings in agreement with the preliminary inclination $i_0$ and $PA_0$ adopted  from the photometric data. In each ring,  circular-rotation model parameters were determined via $\chi^2$-fitting: $PA_{kin}$, the inclination of circular orbits $i$, rotation velocity $V_{rot}$, and systemic velocity $V_{sys}$. We have checked and ruled out the hypothesis that    $V_{sys}$ and $i_{kin}$ change significantly along the radius.  The rotation curve  $V_{rot}(r)$ and the  $PA_{kin}(r)$ distribution are presented in Fig.~\ref{fig:model}. As  follows from this figure, the variations  of $PA_{kin}$ at $r>65''$ are negligible.  This allows us to determine the kinematic orientation parameters of the outer disc as a whole `solid' plane; the corresponding values are listed in   Table \ref{tab:summary}. These values are consistent within the uncertainties with the previous estimates in the literature 
\citep[e.g.,][]{GHASP_IV,Smoker1996}.

\begin{table}
	\centering
	\caption{Derived properties of the NGC~428 gaseous disc}
	\label{tab:summary}
\begin{tabular}{l|c} \hline  \hline \\ [-0.2cm]
Parameter & Value \\
%\MC{1}{c}{ (7) } \\
\\[-0.2cm] \hline \\[-0.2cm]
%RA			& 1:12:55.663   \\
%Dec			& +0:58:53.84	\\
$i_0$ 		& $47\pm5\deg$	\\
$PA_{kin}$	& $118\pm3\deg$ \\ 
$V_{sys}$  & $1149\pm2\kms$ 	\\
%$12+\log(\mathrm{O/H})$ &  8.3?\\
%$O/H_{grad}$ 			&  \\
\\[-0.2cm] \hline \\[-0.2cm]
	\end{tabular}
\end{table}

The final radial distributions  of $V_{rot}$ and   $PA_{kin}$ taken with fixed values of $i$ and $V_{sys}$ are shown in Fig.~\ref{fig:model}. The width of individual rings was $2''$ for the main part of the galaxy and $4''$ for the points with $r>116''$, where only the NW-part of the disc contributes to the corresponding elliptical rings. Also, we assumed that $PA_{kin}=PA_0$ for these outer regions.  

A significant turn of $PA_{kin}$ with a maximum deviation from $PA_0$ of about 20 degrees is observed in the inner part of the gaseous disc ($r<60''$). 

We compared the radial behaviours of $PA_{kin}$ with the results of galaxy isophote fitting using \textsc{iraf}\footnote{\textsc{iraf}: the Image Reduction and Analysis Facility is distributed by the National Optical Astronomy Observatory, which is operated by the Association of Universities for Research in Astronomy, Inc. (AURA) under cooperative agreement with the National Science Foundation (NSF).}, the \textsc{ellipse} task.   The distribution of position angles $PA_{phot}$  along the major axis in the WISE W1-band image is shown in  Fig.~\ref{fig:model}. The analysis of 2MASS and SDSS Stripe82 images was performed in the same way. The results are similar to those obtained for the WISE data, with the exception of those for the innermost part. Here we present only the results of fitting for the WISE data, since they are more reliable and  suffer less from the influence of bright star-forming regions than the SDSS Stripe82 data.

$PA_{phot}$ turns at the same radii but in reverse phase to the $PA_{kin}$ line-of-sight position angle (marked by purple in   Fig.~\ref{fig:model}). According to the discussion above, such behaviour clearly indicates gas motions in a triaxial gravitational potential of the bar. In the paper by \citet{GHASP_IV}, the  authors also claimed that the distortion of central isovelocities  could be treated as a signature of a bar.

It is interesting to note that in the central region (at radii less than $10''$ or $\sim850$ pc), the major axis of NIR isophotes does not coincide with the line of nodes $PA_0$, while $PA_{phot}$ agrees  with $PA_{kin}$ within measurement errors. \citet{Cabrera-Lavers2004} explained the same turn of the nuclear isophotes (together with the peak of their ellipticity $\epsilon$) as the presence of an inner secondary bar. However, the observed alignment 
between the photometric and kinematic PAs disagrees with the gas motions in the inner bar. A more reliable explanation is the presence of a decoupled inner stellar-gaseous disc inclined to the outer disc. Similarly inclined and even polar discs in the circumnuclear regions are well-known structures in both barred and unbarred galaxies   \citep{Corsini2003,Moiseev2012bars}.

The second peculiarity of the velocity is the sharp increase in the rotation velocity by about $15\kms$  at $r>120$ arcsec  that seems inconsistent with the smooth mass distribution in the galactic disc; however, it could be explained by a small $\sim3\deg$ tilt of the disc at these radii. This fact is in agreement with the radio observations by \citet{Smoker1996} who detected a warp in the external \HI disc of NGC 428.

Fig.~\ref{fig:FPI} presents  the model of a quasi-circular velocity field calculated according to the parameters of our tilted-ring model, and the residuals after subtracting our model velocity field from the one observed. Several small-scale regions with relatively high residual velocities of up to $\pm30\kms$ were detected. We discuss their nature below.

\subsection{Small-scale non-circular motions}
\label{sec:resid}

\begin{figure}
\includegraphics[width=\linewidth]{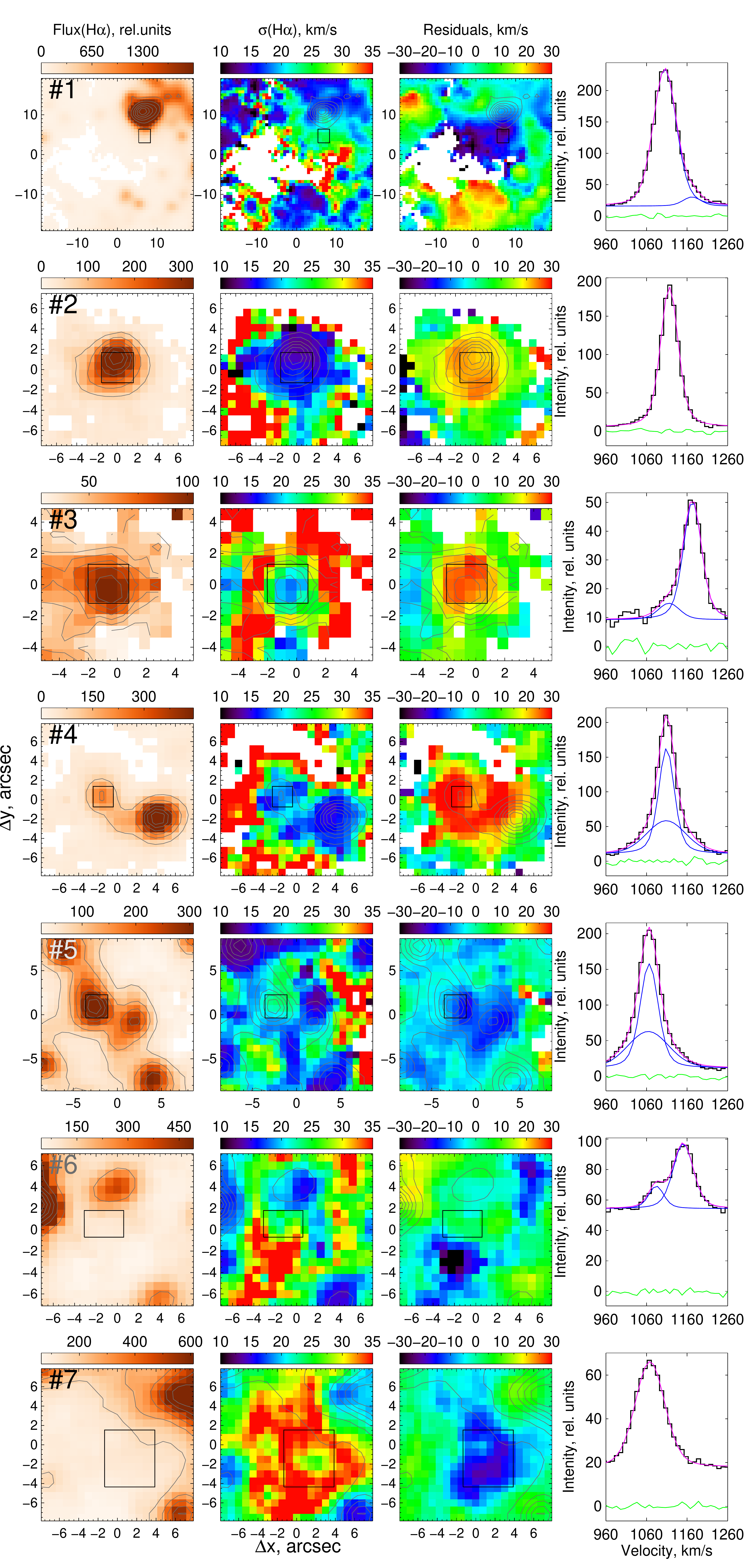}
\caption{Selected regions with high residual velocities, marked in Fig.~\ref{fig:FPI}. From left to right: the H$\alpha$ flux (shown also by contours), velocity dispersion, residual velocity distribution, and an example of the H$\alpha$-line profile (the black curve) integrated over the area shown by the black square together with the result of its decomposition. Each individual component is shown by a blue curve; the magenta curves show the sum of the components; the green curve traces the residuals after model subtraction.}
\label{fig:resid_profile}
\end{figure}

The origin of the residuals in the observed velocity field after subtracting the constructed tilted-ring model (see Fig.~\ref{fig:FPI}) might be related to feedback from winds of massive stars or supernova explosions, or it might highlight the existing gas flows in the area. Here we analyse the H$\alpha$ emission line profile and the distribution of ionized gas in these deviating regions.

Most regions with high residual velocities correspond to the low-brightness area between the nearby bright H~\textsc{ii} regions. Many of them demonstrate a slightly asymmetrical H$\alpha$-line profile or a slightly enhanced velocity dispersion in comparison with the nearby bright regions. All these facts point to slow shock waves from stellar feedback influencing the low-density interstellar medium there. Hence, most of the residual velocities may be explained by the asymmetric or slightly shifted line profiles caused by the local non-circular motions driven by stellar feedback from massive stars. We have selected several such regions that are most intriguing, as well as areas showing high residuals and not related to the low-brightness medium near the H~\textsc{ii} regions. They are marked by circles in Fig.~\ref{fig:FPI}.
Fig.~\ref{fig:resid_profile} shows the H$\alpha$ flux, the velocity dispersion, and the residual velocity distribution together with the H$\alpha$-line profile examples and the results of their multicomponent Voigt decomposition for each of these regions.

Region \#1 exhibits a large area of blue- and redshifted residuals located mostly in a low-density environment between the H~\textsc{ii} regions. However, the brightest H~\textsc{ii} region in the area also partially shows blueshifted residual velocity in its southern part. A redshifted component in the wing of the H$\alpha$-line profile at the edge of the region together with the observed distribution of the velocity dispersion and residuals might be a consequence of the outflow from this H~\textsc{ii} region.

Regions \#2 and \#3 appear as compact H~\textsc{ii} regions with the velocity dispersion distribution typical of H~\textsc{ii} regions and with narrow H$\alpha$-line profiles. A blueshifted H$\alpha$ line component appears in the integrated spectrum of region \#3 with a separation of 60~km~s$^{-1}$. This region probably represents an expanding superbubble in an inhomogeneous medium (e.g., at the former edge of a cloud), which would explain the observed two-component line profile with highly differing intensities of the components. In this case, the expansion velocity of the superbubble should be 30~km~s$^{-1}$, which coincides with the residuals for this region. According to the analytical solution for expanding superbubbles driven by multiple stellar winds and supernova remnants \citep{Weaver1977}, this corresponds to a kinematic age of $t = 0.6 R \,\mathrm{(pc)}/V\,(\mathrm{km\, s^{-1}}) \simeq 3$~Myr, taking into account the 160~pc size of the region. These values are typical for ionized superbubbles around massive OB-associations observed in nearby dwarf galaxies \citep[see, e.g.,][]{Egorov2017, Egorov2018}. In contrast, region \#2 clearly shows a single bright symmetric H$\alpha$-line profile. It seems that the entire region \#2 has a peculiar velocity component with respect to the circular rotation of the disc of the galaxy.

Region \#4 represents two bright H~\textsc{ii} regions connected with each other by a bridge. Most high residual velocities are observed between these H~\textsc{ii} regions and towards the fainter one. While the H$\alpha$-line profile of the bright clump is narrow and symmetrical, it clearly shows a broad underlying component with a velocity dispersion of $\sigma \simeq 43 \, \mathrm{km\, s^{-1}}$ that is four times larger than in its vicinity. 
Similarly, the bright compact region \#5 coinciding with the area of the blueshifted residuals also exhibits a broad ($\sigma \simeq 45 \, \mathrm{km\, s^{-1}}$) underlying component in its H$\alpha$-line profile. The enhanced ratio of [S~\textsc{ii}]/\Ha in this region points to the influence of shock waves (pos. 75~arcsec along the slit PA=$307\degr$, see Section~\ref{sec:spectra}). Some energetic source like a supernova remnant or WR star is likely responsible for the observed ISM turbulence in these two regions.   

Region \#6 corresponds to the nuclear cluster; observations reveal large blueshifted residual velocities and an enhanced \Ha velocity dispersion. However, the most intriguing ionized-gas kinematics are observed between this region and the nearby \HII regions. The
\Ha-line profile there exhibits two narrow components ($\sigma \simeq 12-17 \, \mathrm{km\, s^{-1}}$) with a large separation between them ($67\, \mathrm{km\, s^{-1}}$), and with a high underlying continuum level. Such features might be interpreted as a sign of outflow from the massive nuclear cluster, or as a presence of an expanding superbubble around it. Alternatively, the high velocity motions there might be related to the decoupled inner disc mentioned in Sec.~\ref{sec:kin}.

The most extended area of the enhanced velocity dispersion in the central part of the galaxy (region \#7) also coincides with the blueshifted area on the residual map. The observed H$\alpha$-line profile in individual pixels is slightly asymmetric everywhere in the vicinity of region \#7, but the integral profile is well fitted by a single broad ($\sigma \simeq 31 \, \mathrm{km\, s^{-1}}$) line. We suggest that the enhanced turbulence of the low-density medium is responsible for the observed picture, and one possible explanation is a gas infall. %Alternatively, it could be explained by the shock waves close to the end of the bar.  
% A.M.: No, because the region lies near bar minor axis!

%The region of enhanced velocity dispersion in the centre of the galaxy (see top-right panel of Fig.~\ref{fig:FPI}) also coincides with the blueshifted area on the residuals map. The observed H$\alpha$ line profile is slightly asymmetric there, but it does not clearly separates onto several narrow or broad components. Taking into account the results of performed spectroscopy (see Section~\ref{sec:spectra}), we may conclude that the non-circular motions in this region are caused by the central weak AGN or shock waves in the bar. \red{Carefully check: no, it is not AGN, too far from the center. But there is type Ia SN: SN2013ct nearby. But also too far. Line profile is clearly two-component there, but faint. !!! }

%\red{A.M.: And it's not a bar, because bar's shocks dont' propogate to the very centre of the galaxy (velocity different is too small!!!}

%A few regions showing up on residual map are not related to the low-brightness medium near the H~\textsc{ii} regions as previously mentioned areas. These regions are marked by circles in Fig.~\ref{fig:FPI} \red{to be done}. Further we consider them in more details. Fig.~\ref{fig:resid_profile} demonstrate the H$\alpha$ flux, velocity dispersion and residual velocities distribution together with the examples of H$\alpha$ line profiles and the results of its multicomponent Voigt fitting for each of these regions.

To summarize, most of the residuals between the observed velocity field and the constructed tilted-ring model seem to be related to the regions influenced by stellar feedback from nearby OB associations or individual massive stars. One region (\#2), however, is probably related to some kind of gas flow at the outskirts of the galaxy.%, \red{while the observed high velocity motions in regions \#6 and \#7} 

%Also may obtain the parameters of the companion - magnitude an colours, may estimate the age of stellar population with PEGASE models.

%\begin{figure}
%	\includegraphics[width=\columnwidth]{FIGURES/N428_gri_cut.png}
%    \caption{NGC428}
%    \label{fig:stripe82}
%\end{figure}

\subsection{Gas excitation and chemical abundance}
\label{sec:spectra}

\begin{figure}
	\includegraphics[width=\columnwidth]{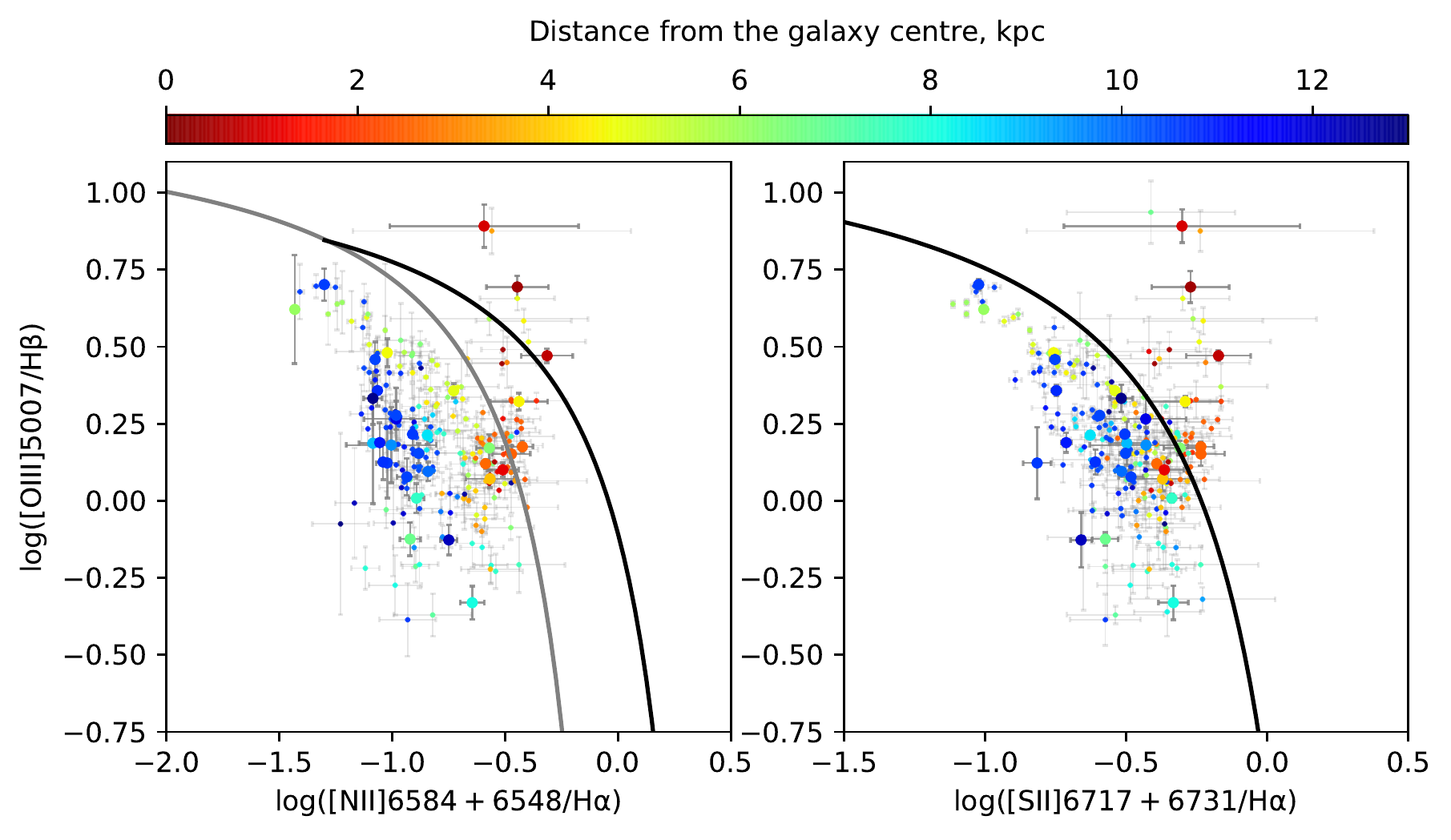}
    \caption{Diagnostic diagrams \OIIIHb\, vs \NIIHa\, (left-hand panel) and \SIIHa\, (right-hand panel) constructed for the integrated spectra of individual \HII regions (large circles) and for individual binned pixels (small circles) in NGC~428. Different symbol colours correspond to different radial distances.
The black curve in both panels represents the `maximum starburst line' from \citet{Kewley2001}, and the grey curve from \citet{Kauffmann2003} in the left-hand panel separates the pure star-forming regions from those with a composite mechanism of excitation.}
    \label{fig:BPT}
\end{figure}

\begin{figure}
	\includegraphics[width=\columnwidth]{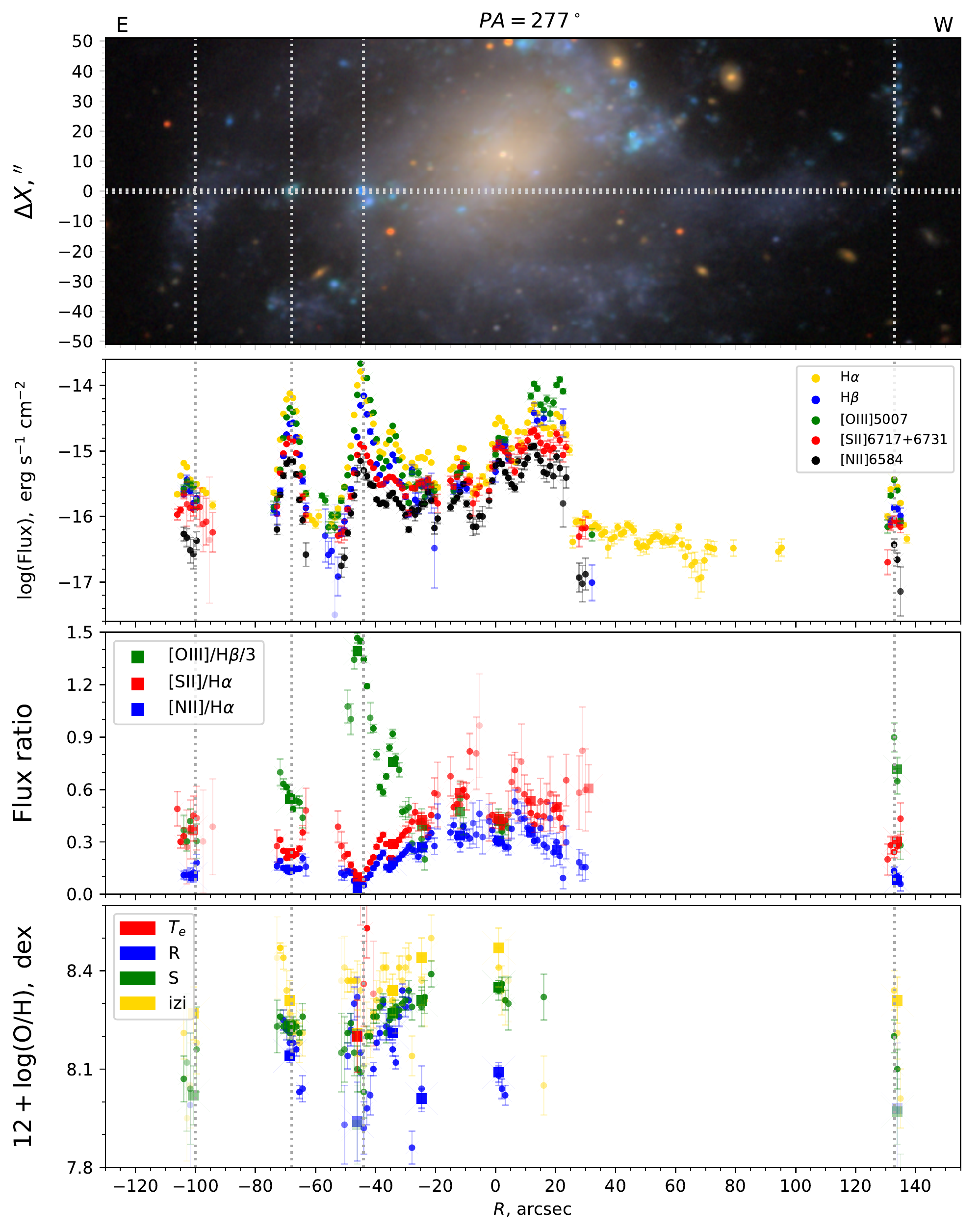}
    \caption{Distribution of the emission line fluxes (the top panel), their ratios (the middle panel) and oxygen abundance derived using several methods (the bottom panel) along the slit PA=277. The horizontal dashed line in the top panel shows the position of the slit overlaid on the composite SDSS-$i$,$r$,$g$ image of the NGC428 galaxy. The squares in the two bottom panels correspond to the values obtained from the integrated spectra of individual \HII regions, while the circles show the values derived for each binned pixel along the slit. For clarity, the transparency of the symbols correlates with the corresponding uncertainty of the measured metallicity. All the points corresponding to the non-photoionization mechanism of excitation according to Fig.~\ref{fig:BPT} are masked.}
    \label{fig:Chem1}
\end{figure}

\begin{figure}
	\includegraphics[width=\columnwidth]{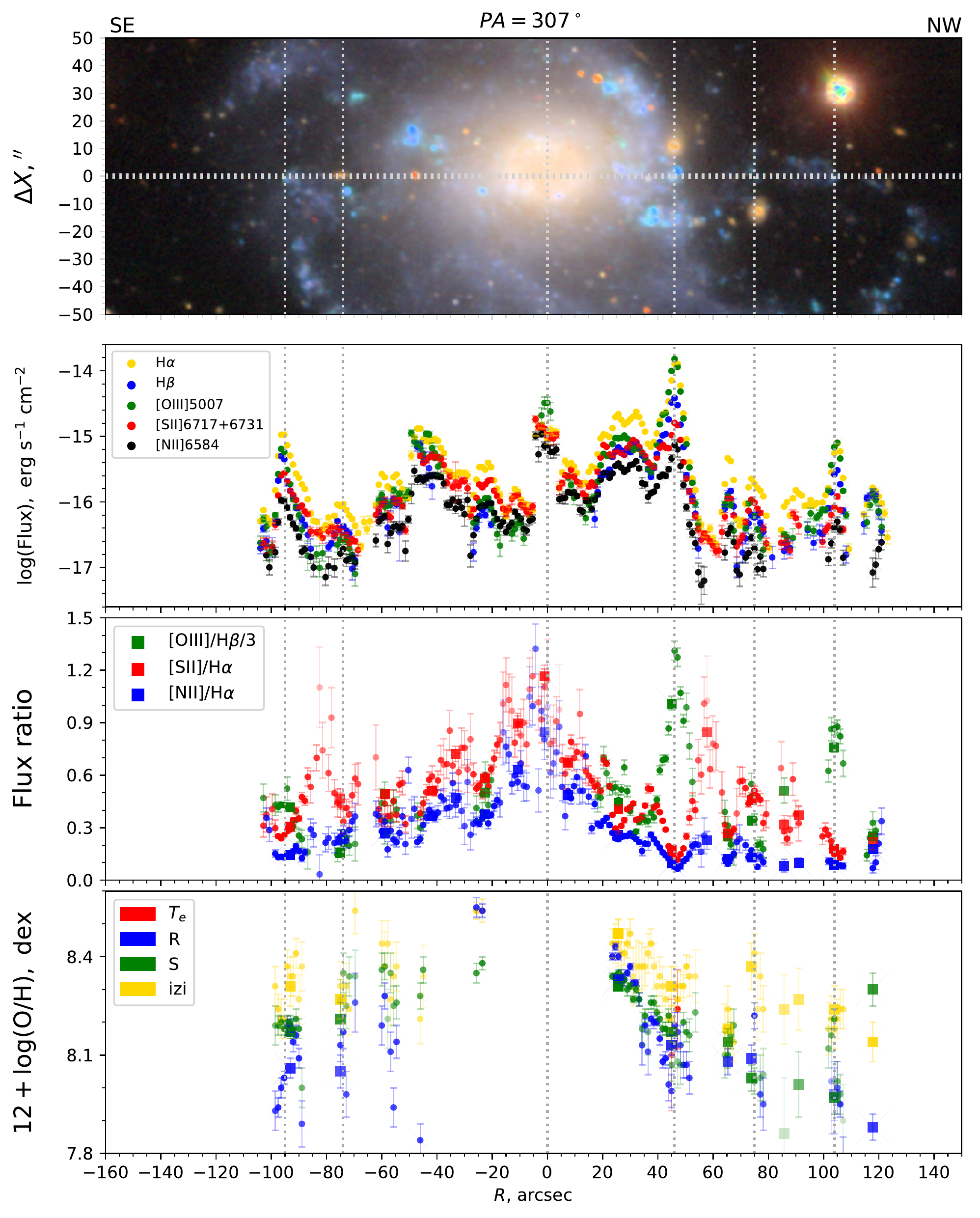}
    \caption{Same as Fig.~\ref{fig:Chem1} but for the slit PA=307.}
    \label{fig:Chem2}
\end{figure}

\begin{figure}
	\includegraphics[width=\columnwidth]{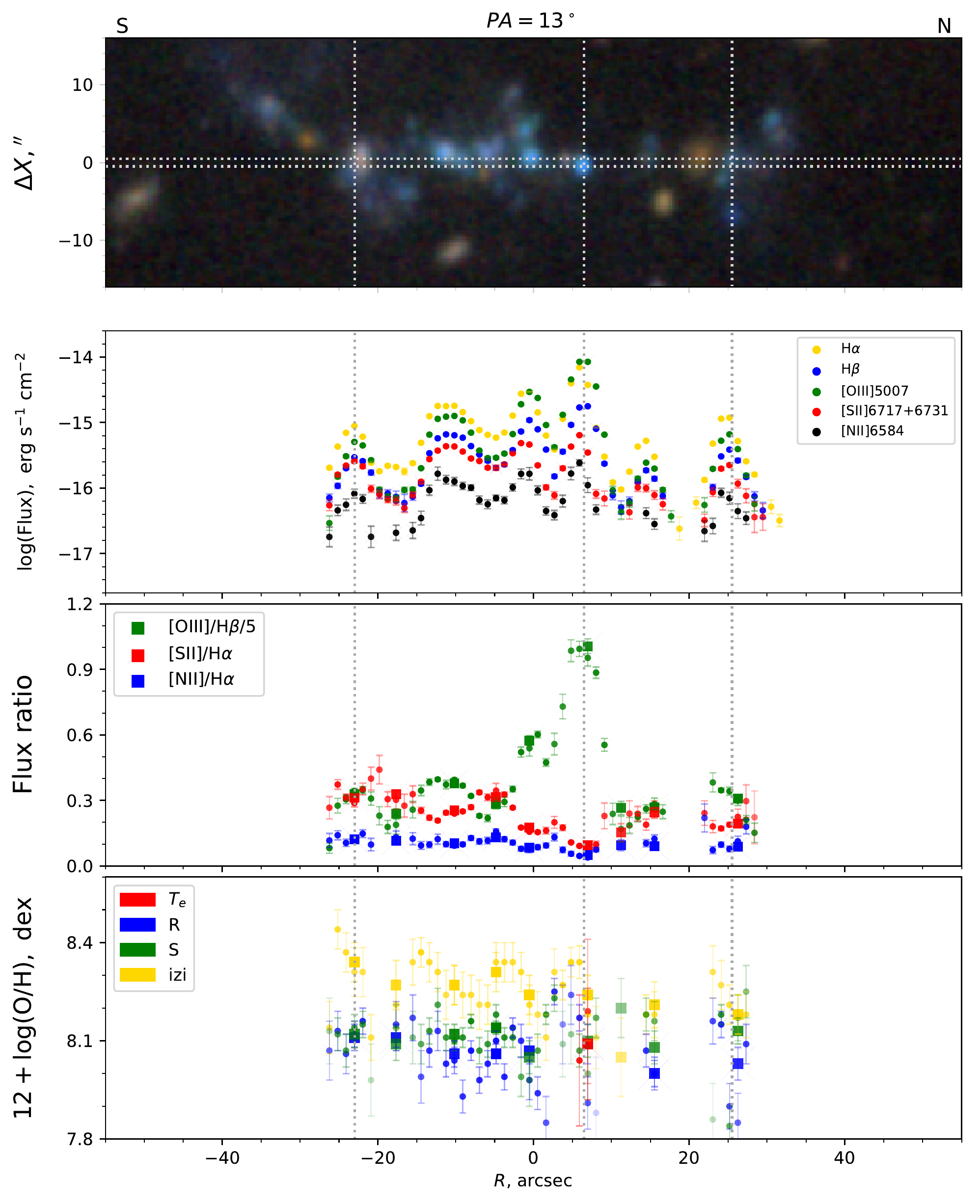}
    \caption{Same as Fig.~\ref{fig:Chem1} but for the slit PA=13.}
    \label{fig:Chem3}
\end{figure}

Long-slit spectra were used to analyse the mechanism of excitation of the emission from different regions in the galaxy, as well as for deriving its oxygen abundance $\mathrm{12+\log(O/H)}$, which is an indicator of gas metallicity. We analysed the emission-line ratios along each slit. In order to increase the signal-to-noise ratio, we also studied the integrated spectra of individual \HII regions or extended diffuse areas intersected by the slit (the results obtained for such regions are shown by large black circles in Fig.~\ref{fig:BPT} and by squares in Figs.~\ref{fig:Chem1}--\ref{fig:Chem3}).

The positions of the regions crossed by the slits in a classical BPT \citep*{BPT} diagnostic diagram are shown in Fig~\ref{fig:BPT}. The `maximum starburst line' \citep{Kewley2001} demarcation separates the regions whose emission can be explained by photoionization by young massive stars as a consequence of ongoing star formation from those with major contribution from other sources of excitation (e.g., shocks). The points lying between the \citet{Kewley2001} curve and the \citet{Kauffmann2003} curve (grey in Fig. \ref{fig:BPT}) correspond to a composite excitation mechanism.
%Those points lying between that line and another one by \citet{Kauffmann2003} correspond to the composite mechanism of excitation. 
Such regions are probably ionized by harder radiation (like diffuse ionized gas, see \citealt{Zhang2017}), or they may be subjected to an enhanced influence of shock waves. A photo-ionising mechanism of excitation is evident from the diagram for all the regions at a galactocentric distance of $R>3$~kpc except one. A composite mechanism of excitation is observed towards the centre of the galaxy; shock excitation should play a significant role there. The shocks might be produced  by collisions between gaseous clouds in the circumnuclear inclined disc and the gas in the main galactic plane (Section~\ref{discussion}). A few points lying inside the central 1 kpc fall above the `maximum starburst line' and are indicative of the AGN activity there.

As follows from Figs.~\ref{fig:Chem1}--\ref{fig:Chem3}, most regions outside the central part demonstrate line flux ratios typical of \HII regions. Nevertheless, an enhanced \SIIHa\, flux ratio is observed in a few non-central regions (positions $-90$, 60 and 75 arcsec in Fig.~\ref{fig:Chem2}). The first two regions correspond to the star-forming complexes near the  ends of the bar, i.e., regions where the gas is accumulated under the influence of the non-axisymmetric potential of the bar \citep[][and references therein]{Renaud2015}.     
However, the residual velocity map reveals no kinematic  peculiarities there, because the line-of-sight projection of possible radial streaming motions is negligible near the major axis of the disc. The third region corresponds to the \HII region \#5 with an underlying broad component (see Section~\ref{sec:resid}). 

Unfortunately, a weak [O~\textsc{iii}] 4363 \AA\ emission line sensitive to electron temperature was confidently detected for only two bright regions in our spectra. Because of this we were unable to use the $T_e$ method for measuring the oxygen abundance in most of the H~\textsc{ii} regions. Instead, we selected several methods from the widely used empirical `strong emission line' (SEL) techniques: the $R$- and $S$-methods from \cite{Pilyugin2016}, and the \textsc{izi} method from \cite{izi} with the photoionization model from \citet{Levesque2010}. These methods were selected because they provide reliable estimates in the NGC~428 metallicity range; the first two also yield results consistent with the $T_e$ method, while for the \textsc{izi} method a systematic offset of up to $\sim 0.1$ dex is expected due to the well known discrepancy problem between the empirical and theoretical SEL methods \citep[see, e.g.,][]{Kewley2008,Lopez-Sanchez2012}. All these methods require different sets of observed flux ratios:
the $R$-method is based on the ratios of [O~\textsc{ii}] 3727, 3729~\AA\,, [N~\textsc{ii}] 6584~\AA\,, and \OIII 5007~\AA\, to H$\beta$; the $S$-method requires the [S~\textsc{ii}] 6717, 6731~\AA\, line fluxes instead of [O~\textsc{ii}]; the \textsc{izi} method based on photoionization models uses all the available emission line data.

Because all the methods used for estimating the metallicity are calibrated by \HII regions or pure photoionization models, we have excluded all the regions lying above the `maximum starburst line' or in the composite zone from further analysis.
The distributions of oxygen abundance along the slits, calculated using each method, are shown in Figs.~\ref{fig:Chem1}--\ref{fig:Chem3}. All the calibrations used show similar metallicity trends along the slits but with an expected systematic offset. Significant deviations of the results obtained with the $R$-method from those derived using other methods in the centre of the galaxy are probably caused by an uncertainty of the [O~\textsc{ii}]/H$\beta$ ratio due to significant extinction variations and enhancement: the colour excess there (derived from the observed Balmer decrement) is $E(B-V) = 0.5 - 1.3$ mag, while the values $E(B-V) = 0.07 - 0.3$ mag are typical for other parts of the galaxy. The estimates derived by the empirical $S$-method are in very good agreement with those obtained with the $T_e$-method for the two [O~\textsc{iii}]-bright regions.

\begin{table*}
\caption{Oxygen abundance and its gradient measured for NGC~428 using different methods}
\label{tab:oh}
%\begin{tabular}{lcccc}
%\hline
%Method & $T_e^*$ & $R$ & $S$ & $izi$ \\
%\hline
%$\mathrm{12+\log(O/H)_0}$, dex & $8.35\pm0.35$ & $8.23\pm0.06$  & $8.33\pm0.05$ & $8.46\pm0.04$ \\
%$\mathrm{12+\log(O/H)_{R_{25}/2}}$, dex & $8.22\pm0.15$ & $8.13\pm0.07$ & $8.22\pm0.06$& $8.36\pm0.04$\\
%$\Delta\mathrm{(O/H),\ dex\ kpc}^{-1}$ & $-0.024\pm0.045$ &$-0.019\pm0.007$ & $-0.022\pm0.006$ & $-0.020\pm0.004$\\
%$\Delta\mathrm{(O/H),\ dex}\ R_{25}^{-1}$ & & & & \\
%$\Delta\mathrm{(O/H),\ dex}\ R_e^{-1}$ & & & & \\
%\hline
%\end{tabular}

\begin{tabular}{lccccc}
\hline
Method & $\mathrm{12+\log(O/H)_0}$, dex  & $\mathrm{12+\log(O/H)_{R_{25}/2}}$, dex & $\Delta\mathrm{(O/H),\ dex\ kpc}^{-1}$ & $\Delta\mathrm{(O/H),\ dex}\ R_{25}^{-1}$ & $\Delta\mathrm{(O/H),\ dex}\ R_e^{-1}$ \\
\hline
$T_e^*$ & $8.35\pm0.35$ &$8.22\pm0.15$ & $-0.024\pm0.045$ & $-0.25\pm0.47$ & $-0.09\pm0.17$\\
$R$  &  $8.23\pm0.06$  & $8.13\pm0.07$ & $-0.019\pm0.007$&$-0.20\pm0.07$  & $-0.07\pm0.03$ \\
$S$ & $8.33\pm0.05$ &$8.22\pm0.06$ & $-0.022\pm0.006$ &$-0.22\pm0.06$  &$-0.08\pm0.02$  \\
\textsc{izi} & $8.46\pm0.04$&$8.36\pm0.04$ &$-0.020\pm0.004$ & $-0.21\pm0.04$ & $-0.08\pm0.02$\\
\hline
\end{tabular}
\begin{tablenotes}
\item $^*$ Because $T_e$ measurements are available for only two regions, the uncertainties of the derived values were calculated using the observed uncertainties for those two points
\end{tablenotes}
\end{table*}

\begin{figure}
	\includegraphics[width=\columnwidth]{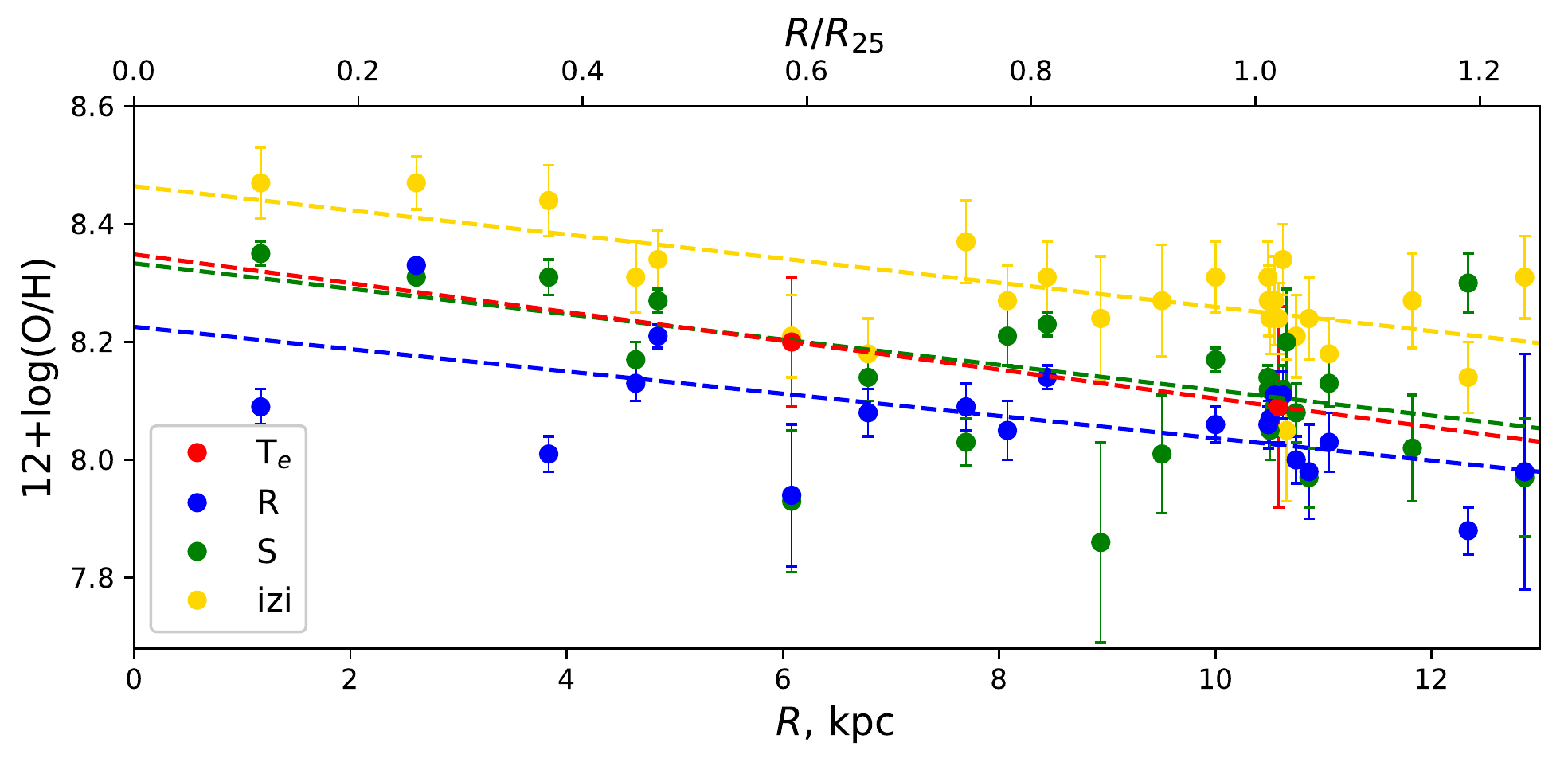}
    \caption{Radial distribution of oxygen abundance in the galaxy NGC~428 derived by four methods. The dashed coloured lines show the results of linear least-squares fitting.} 
    \label{fig:chemrad}
\end{figure}

The oxygen abundance is almost uniform in the outer part of the western star-forming ring (see Fig.~\ref{fig:Chem3}), while it shows a gradient towards the centre (see Figs.~\ref{fig:Chem1}--\ref{fig:Chem2}). In order to investigate the presence of a metallicity gradient, in Fig.~\ref{fig:chemrad} we plot all measurements of $12+\log(\mathrm{O/H})$ along all the slits over the deprojected galactocentric distance of the region. For deprojection, we used the inclination and position angles of the galactic disc derived in Section~\ref{sec:kin}. Despite the large spread of individual measurements, all the methods used reveal a detectable and very similar metallicity gradient of about $-0.02\ \mathrm{dex\ kpc^{-1}}$, or $-0.21\ \mathrm{dex}\ R_{25}^{-1}$ ($R_{25}=122''\approx10.5$ kpc according the NED).

The oxygen abundance in NGC~428 was previously studied by \citet{Pilyugin2014} and \citet{Eridanus} from SDSS spectra. \citet{Pilyugin2014} found a flat metallicity distribution with the central value  $12+\mathrm{\log(O/H)}=8.20\pm0.06$, however, this estimate is based on a small number of regions concentrated mainly at the same distance from the galaxy centre. The value $\mathrm{12+\log(O/H)}=8.12\pm0.06$ estimated in \citet{Eridanus} and obtained for one bright \HII region is consistent with the values from \citet{Pilyugin2014} within the uncertainties. Our estimate of the oxygen abundance at the radius $R=0.5R_{25}$ is consistent with those obtained in the above works, and a much better completeness in terms of galactocentric radii allowed us to measure the metallicity gradients (see Table~\ref{tab:oh}).

%In paper by \citet{Pilyugin2014} the ``C-method'' by \citet{Pilyugin2012} was used to estimate the metallicity gradient for NGC~428 using SDSS spectral data. The gradient was found to be flat with central metallicity $12+log(O/H)_{R0}=8.20\pm0.06$. Though it should be notes that this estimation is based on the small number of regions concentrated mainly on the same distance from the galaxy centre.
%\textbf{The value $12+log(O/H)=8.12\pm0.06$ estimated in Kniazev et al, 2018 (submitted to MNRAS) is consistent with values from \citet{Pilyugin2014} within the errors. It was obtained for one bright HII region using the SDSS spectral data with the modified Te method \citep{Kniazev2003,Kniazev2004}. Our estimate is consistent with values obtained in \citet{Pilyugin2014} and Kniazev et al, 2018 (submitted).}

\section{Discussion}
\label{discussion}

Optical images of NGC 428 demonstrate a very peculiar morphology with several large-scale asymmetric structures that are possibly tidal or post-interaction features. In contrast to this perturbed picture, the large-scale kinematics of ionized gas seem to be `quiet'. Indeed,  if the radial streaming motions in the bar are taken into account at galactocentric distances of 1--4 kpc, the velocity field is well described by circular rotation of a thin flat disc. Circular rotation dominates even in the area of the giant offset star-forming ring at the western outskirts of the region. In this region, we detected only a  very small tilt of the gaseous-cloud orbits by a few degrees relative to the disc inclination inside the optical radii $R_{25}$. However, a careful analysis of our new deep \Ha velocity field allowed us to detect  two kinematic  peculiarities  in the gaseous disc of NGC 428, which could be related to some external event:

\begin{itemize}
\item[(i)] An inclined disc in the circumnuclear (at $r<10''=0.85$ kpc) regions.
\item[(ii)] An outer \HII region \#2 with significant residual velocities.
\end{itemize}

Reviews of the properties of the inner decoupled discs are presented by \citet{Corsini2003} and \citet{Moiseev2012bars}. These authors argued that the origin of the majority of such structures is closely related to the capture of external matter having a spin different from that of the host galaxy. For instance, about 2/3  of the galaxies in the catalogue of inner polar discs and rings reveal various signs of recent interaction or merging \citep{Moiseev2012bars}. A good illustration is NGC 7217, for which \citet{Sil2011} demonstrated an agreement of the observed properties of the galaxy with the results of the GalMer database simulations of a wet minor merger with a gas-rich dwarf companion. 

If we adopt $PA=105\pm5\degr$ and $\epsilon=0.2\pm0.05$ for this inner disc in NGC 428, then the mutual angle between the inner and outer discs is either $13\pm7\degr$ or $83\pm8\degr$ according to eq.(1) in \citet{Moiseev2012bars}. The present ambiguity with these two solutions  is  caused by the fact that $PA$ and $i$ are insufficient to describe the full orientation of a disc in space --- we should also know which half of the disc is nearest to the observer. If the spiral structure in NGC 428 is trailing, we can conclude that the SW half of the outer disc corresponds to the nearest  side of the galaxy. At the same time, the  orientation of the  inner disc relative the external one is still unknown.

If we believe the first solution for the mutual angle between the discs, then this slightly inclined disc should be dynamically unstable against precession within a few dynamic timescales of the inner regions ($<100$ Myr). On the other hand, the stability of polar and inclined circumnuclear structures is a matter of debate in contrast to the well-studied case of large-scale polar rings \citep{Arnaboldi1994}.

An additional hint on the existence of an inclined nuclear disc is the distribution of gas excitation properties along $PA=277\degr$, which is near the major axis of this possible disc (Fig.~\ref{fig:Chem1}). The figure shows that the [N~\textsc{ii}]/\Ha- and [S~\textsc{ii}]/\Ha-line ratios have peaks at a distance of $\sim10$ arcsec on both sides of the nucleus. This sign of shock excitation may be related to the direct interaction between gaseous clouds in the outer and inner discs. A similar picture was observed, for instance, in Arp 212 \citep{Moiseev2008arp}.

The second kinematically decoupled region in NGC~428 is the HII knot \#2 (see Section~\ref{sec:resid} and Figs.~\ref{fig:FPI}, \ref{fig:resid_profile}). This bright compact region demonstrates a narrow unperturbed \Ha-line profile with the line-of-sight velocity significantly differing from that predicted by the galaxy rotation model. A flow of gas caused by external accretion might explain such ionized-gas kinematics behaviour at this location. Unfortunately, we have no chemical abundance estimates for this region.

The estimated oxygen abundance gradient for NGC~428 ($-0.21\ \mathrm{dex}\ R_{25}^{-1}$, or $-0.08\ \mathrm{dex}\ R_{e}^{-1}$, taking the effective radius $R_e = 44.68$ arcsec from \citealt{Vika2013}) agrees well with the median value obtained for a large sample of MANGA galaxies ($\sim -0.08$ dex/$R_{e}$ according to \citealt{Belfiore2017}) but is lower than the estimates derived from the CALIFA data ($\sim -0.11$ dex $R_{e}^{-1}$, according to \citealt{Sanchez2014}). \cite{Ho2015} obtained a median metallicity gradient of $\sim -0.39$ dex/$R_{25}$ for a large sample of local star-forming galaxies, which is also significantly higher than our estimate for NGC~428. \citet{Bresolin2015} have showed that there is a linear metallicity gradient trend with $1/R_d$, where $R_d = R_e/1.6783$ is the disc scalelength. From their fig. 3, we have found the expected value of the metallicity gradient for NGC~428 to be $\sim -0.04$ dex kpc$^{-1}$, which is twice greater than the one observed. 
%However the used value of $R_e$ was derived without separation onto bulge, bar and disc components. \citet{Cabrera-Lavers2004} performed such photometric decomposition and obtained a flat profile for the disc of NGC~428. This lead to significant increase of the $R_d$ value, and hence to decrease of the expected metallicity gradient.

Together with the existence of an inclined circumnuclear disc and \HII region \#2 decoupled from pure circular rotation, the observed diminished metallicity gradient points to a probable recent episode of  metal-poor gas accretion (minor merging with a gas-rich satellite or a capture of external gas clouds) onto the disc of the galaxy. This event might flatten the observed gradient and produce the observed features in the morphology and kinematics.

%The NGC 428 was previously included in GHASP (Gassendi $H\alpha$ survey of SPirals) survey \citep{GHASP0,GHASP_I}. The observations of NGC 428 with Fabry-Perot interferometer were performed in the frame of this survey, and velocity field in $H\alpha$ hydrogen line was derived. % (with a sampling about 5 km s$^{−1}$ in velocity and 3 arcsec in spatial resolution). 
%In the paper by \citet{GHASP_IV} authors found that the velocity field in the central part shows distortions that could be treated as a signature of a bar.
%Also the ionized gas kinematics in $H\alpha$ line obtained with GH$\alpha$FaS was studied in \citet{ErrozFerrer2015} and non-circular motions in central part were measured for NGC 428.
%  Там просто ужасное поле, недостойное рассмотрения!!

\section{Conclusions}
\label{conclusions}

We describe our project aimed at the study of interactions and gas accretion in the void environment. A sample of  late-type galaxies was selected according to their reduced metallicity (in comparison with the reference 'metallicity-luminocity' relation) and/or signs of disturbances in the optical morphology.  Ten of eighteen objects are strongly isolated galaxies, while the others are mainly members of groups and pairs. The observed data involved in the project includes: long-slit spectroscopy obtained for the study of the chemical abundance distributions in the discs and outskirts of the galaxies,  3D-spectroscopy with a scanning FPI (mapping the ionized-gas kinematics), and archival optical (SDSS) and IR (WISE, 2MASS) images for surface photometry and isophotal analysis.  The long-slit and scanning FPI data were collected with the   SCORPIO/SCOPRIO-2 focal reducers at the 6-m Russian telescope of SAO RAS.

Observations of the brightest target in the sample, the SABm galaxy NGC~428, are considered as an illustration of our methods, data reduction, and analysis algorithms. At first glance, the galaxy looks like an ordinary Magellanic-type spiral galaxy with a more or less uniform radial distribution of the chemical abundance. Large-scale gas kinematics is dominated by pure circular rotation with radial streaming motions in the bar region. Some local (on a spatial scale of 0.5-1 kpc) perturbations of the `quiet' ionized-gas kinematics (the double-component \Ha emission line profiles, enhanced residual velocity,  and velocity dispersion) are unambiguously  related to stellar feedback in the regions of violent star formation. 

However, a more careful analysis of the collected data reveals several interesting features:

\begin{enumerate}
\item A circumnuclear ($r<850$ pc) disc nested in the large-scale bar and inclined at $13\pm7\degr$ or $83\pm8\degr$ to the main galaxy plane.

\item  \HII region \#2 with a narrow one-component emission line profile and significantly large residual velocities.

\item The oxygen abundance gradient  seems to be slightly lower than that expected for similar local galaxies. 
\end{enumerate}

Individually, these features can not be considered as strong evidence of an external event in the history of NGC 428. However, all of them combined, and also the perturbed and asymmetric morphology in the outer region of the galaxy, suggest a gas accretion or minor merging episode in the recent history of NGC 428 on a timescale of $\sim0.5$~Gyr or less (a few revolutions of the outer galaxy disc).

%\red{In the course of this project, we have found several unusual objects with very strong misalignments between their morphology and kinematics. The results of their study will be presented in  upcoming papers.}

\section*{Acknowledgements}

We thank the anonymous referee for detailed comments that helped us to improve this manuscript.
This work was supported by the Russian Foundation for Basic Research (project no. 17-32-50013). Observations with the 6-m Russian telescope (BTA) were carried out with the financial support of the Ministry of Education and Science of the Russian Federation (agreement no. 14.619.21.0004, project ID RFMEFI61914X0004). The authors thank Olga Sil'chenko for the help and inspiration for this work, and Roman Uklein and Dmitry Oparin for their assistance in observations with the 6-m telescope.
This publication makes use of the data products from the Wide-field Infrared Survey Explorer, which is a joint project of the University of California, Los Angeles and the Jet Propulsion Laboratory/California Institute of Technology, and NEOWISE, which is a project of the Jet Propulsion Laboratory/California Institute of Technology. WISE and NEOWISE are funded by the National Aeronautics and Space Administration. This research has made use of the NASA/IPAC Extragalactic Database (NED) which is operated by the Jet Propulsion Laboratory, California Institute of Technology, under the contract with the National
Aeronautics and Space Administration.
The authors acknowledge the spectral and photometric data and the related
information available in the SDSS database used for this study.
Funding for SDSS-III has been provided by the Alfred P. Sloan Foundation,
the Participating Institutions, the National Science Foundation, and the
U.S. Department of Energy Office of Science. The SDSS-III web site is
http://www.sdss3.org/.
SDSS-III is managed by the Astrophysical Research Consortium for the
Participating Institutions of the SDSS-III Collaboration including
the University of Arizona, the Brazilian Participation Group, Brookhaven
National Laboratory, Carnegie Mellon University, University of Florida,
the French Participation Group, the German Participation Group, Harvard
University, the Instituto de Astrofisica de Canarias, the Michigan
State/Notre Dame/JINA Participation Group, Johns Hopkins University,
Lawrence Berkeley National Laboratory, Max Planck Institute for Astrophysics,
Max Planck Institute for Extraterrestrial Physics, New Mexico State
University, New York University, Ohio State University, Pennsylvania
State University, University of Portsmouth, Princeton University, the
Spanish Participation Group, University of Tokyo, University of Utah,
Vanderbilt University, University of Virginia, University of
Washington, and Yale University.

%%%%%%%%%%%%%%%%%%%%%%%%%%%%%%%%%%%%%%%%%%%%%%%%%%



\bibliographystyle{mnras}
\bibliography{accretion}

%%%%%%%%%%%%%%%%%%%%%%%%%%%%%%%%%%%%%%%%%%%%%%%%%%

% Don't change these lines
\bsp	% typesetting comment
\label{lastpage}
\end{document}